\documentclass[11pt,reqno]{amsart}
\usepackage{amsmath}
\usepackage{amssymb}
\usepackage{amsthm}
\usepackage{amscd, amsfonts, mathrsfs}

\usepackage{cases}

\usepackage[dvips]{epsfig}
\usepackage{verbatim} 

\usepackage{epsf}
\usepackage[bookmarksnumbered,pdfpagelabels=true,plainpages=false,colorlinks=true,
            linkcolor=black,citecolor=black,urlcolor=black]{hyperref}

\theoremstyle{plain}
\newtheorem{theorem}{Theorem}[section]
\newtheorem{claim}[theorem]{Claim}

\newtheorem{proposition}{Proposition}[section]

\newtheorem{assumption}[theorem]{Assumption}

\theoremstyle{definition}
\newtheorem{remark}[theorem]{Remark}
\newtheorem{notation}[theorem]{Notation}

\newtheorem{convention}[theorem]{Convention}

\numberwithin{equation}{section}
\allowdisplaybreaks

\newcommand{\cL}{\mathcal L}

\newcommand{\al}{\alpha}
\newcommand{\be}{\beta}
\newcommand{\ga}{\gamma}
\newcommand{\Ga}{\Gamma}
\newcommand{\de}{\delta}

\newcommand{\la}{\lambda}
\newcommand{\La}{\Lambda}
\newcommand{\si}{\sigma}
\newcommand{\Si}{\Sigma}

\newcommand{\Om}{\Omega}

\newcommand{\RR}{\mathbb R}

\newcommand{\rar}{\rightarrow}

\newcommand{\ve}{\varepsilon}

\newcommand{\mss}{\hspace{0.2cm}}

\newcommand{\sv}{\vartheta} 
\newcommand{\bv}{\zeta} 
\newcommand{\hc}{\kappa} 

\newcommand{\st}{\mathsf{M}}

\title[First order]{On a viable first order formulation of
relativistic viscous fluids and its applications to cosmology}

\author[Disconzi]{Marcelo M. Disconzi}
\address{Department of Mathematics\\
Vanderbilt University\\ Nashville, TN, USA}
\email{marcelo.disconzi@vanderbilt.edu}

\author[Kephart]{Thomas W. Kephart}
\address{Department of Physics and Astronomy\\ Vanderbilt University\\
Nashville, TN, USA}
\email{thomas.w.kephart@vanderbilt.edu}

\author[Scherrer]{Robert J. Scherrer}
\address{Department of Physics and Astronomy\\ Vanderbilt University\\
Nashville, TN, USA}
\email{robert.scherrer@vanderbilt.edu}

\begin{document}

\maketitle

\begin{abstract}
We consider a first order formulation of relativistic fluids with bulk viscosity
based on a stress-energy tensor introduced by Lichnerowicz. 
Choosing a barotropic equation of state, we show that this theory satisfies basic
physical requirements and, under the further assumption of vanishing vorticity, that the 
equations of motion are causal, both in the case of a fixed background and when the equations
are coupled to Einstein's equations. Furthermore, Lichnerowicz's proposal does not fit into
 the general framework
of first order theories studied by Hiscock and Lindblom, and hence their instability results do not
apply. These conclusions apply to the full-fledged non-linear theory, without any equilibrium or near
equilibrium assumptions. 
Similarities and differences between the approach explored here and other 
theories of relativistic viscosity,  including the Mueller-Israel-Stewart formulation, are addressed. 
Cosmological models based on the Lichnerowicz stress-energy tensor are studied.
As the topic of (relativistic) viscous fluids is also of interest outside the general relativity and cosmology communities, 
such as, for instance, in applications involving heavy-ion collisions, we 
make our presentation is largely self-contained.
\end{abstract}

\newpage

\tableofcontents

\part{INTRODUCTION AND STATEMENT OF THE RESULTS}

\section{Introduction\label{section_intro}}

The coupling of matter to gravity in general relativity is usually accomplish by specifying
a stress-energy tensor for the matter fields, which is then sourced to Einstein's equations. When the matter fields
admit a Lagrangian formulation for its equations of motion, there exists a well-defined
procedure that uniquely determines the corresponding stress-energy tensor in the context
of general relativity. Such a Lagrangian formulation, in turn, rests upon 
a variational formulation of the corresponding non-relativistic equations of motion,
see, e.g.,  \cite{HawkingEllisBook, WaldBookGR1984} for details.
This procedure fails if one wishes to describe the dynamics of a fluid with viscosity interacting with gravity
in that the classical Navier-Stokes equations are not known to be the Euler-Lagrange equations
of a specific action functional. While this does not prevent us from writing a stress energy-tensor 
for the Navier-Stokes equations, its generalization to general relativity is not unique, and, in fact, presents
ambiguities \cite{Weinberg_GR_book}. A review of some of the
different proposals for handling viscosity in relativity is given in section
\ref{section_theories}. Section \ref{section_theories} also discusses similarities and differences
between the the approach studied in this
paper and other theories of relativistic viscosity.

In this work we shall take the following stress-energy tensor
to describe relativistic viscous fluids:
\begin{align}
\begin{split}
T_{\al\be} & = (p + \varrho) u_\al u_\be + p g_{\al\be} 
-(\bv - \frac{2}{3} \sv) \pi_{\al\be} \nabla_\mu C^\mu  - \sv \pi_\al^\mu \pi_\be^\nu
(\nabla_\mu C_\nu + \nabla_\nu C_\mu) \\
&  - \hc (q_\al C_\be + q_\be C_\al) + 2\sv \pi_{\al\be} u^\mu \nabla_\mu F.
\end{split}
\label{Lichnerowicz_stress}
\end{align}
$p$ and $\varrho$ are scalar functions representing the pressure and (total) energy density of the fluid;
$g$ is a Lorentzian metric;  $\nabla$ is the covariant derivative (Levi-Civita connection) associated with $g$;
$u_\al$ is the four-velocity\footnote{We adopt the Eckart frame; see section \ref{section_velocity_fluid}.} of fluid particles, which satisfies
\begin{gather}
u^\al u_\al = -1;
\label{normalization_u}
\end{gather}
$\sv$ and $\bv$ are the coefficients of shear and bulk viscosity, respectively, and
$\kappa$ is the coefficient of thermal conductivity. They are assumed to be non-negative functions of the thermodynamic
variables (i.e., of  $\varrho$, $p$, and possibly other thermodynamic quantities 
introduced below, 
such as the rest mass density, temperature, entropy, and internal energy).
Their particular form depends on the nature of the fluid. 
$q_\al$ is the fluid's heat flux whose specific form will not be important here.
\begin{gather}
\pi_{\al\be} = g_{\al\be} + u_\al u_\be
\label{Projector}
\end{gather}
is the projection 
onto the space orthogonal to $u$, i.e., 
\begin{gather}
\pi_{\al\be} u^\be = 0.
\label{projection_u}
\end{gather}
$C_\al$ is the dynamic velocity of the fluid, also known as the enthalpy current,
and defined by 
\begin{gather}
C_\al = F u_\al, 
\label{def_C}
\end{gather}
where $F$ is the specific enthalpy of the fluid, 
given by 
\begin{gather}
F = \frac{p + \varrho}{n},
\label{def_enthalpy}
\end{gather}
where $n$ is the rest mass density, which satisfies
\begin{gather}
\nabla_\mu (n u^\mu ) = 0,
\label{rest_mass_eq}
\end{gather}
as long as particle number is conserved.  When particle number is not conserved,
Eq. (\ref{rest_mass_eq}) must be modified to
\begin{gather}
\nabla_\mu (n u^\mu ) = \sigma,
\label{rest_mass_not_conserved}
\end{gather}
where $\sigma$ is the net number of particles created per unit time per unit volume
in a co-moving fluid element.  The particular form of $\sigma$ will depend on the processes involved and
should ultimately be derived from microscopic kinetic theory.
From a macroscopic point of view, $\si$ is considered a given known function.  However, regardless of whether or not $\sigma = 0$,
we can always define $F$ to be given by Eq. (\ref{def_enthalpy}). Most of the discussion
in this paper focuses on $\si = 0$.

Naturally, all of the above functions and tensors are defined on 
a four-dimensional differentiable manifold, the space-time, henceforth denoted by 
$\st$. In the above, we have adopted:
\begin{convention}
the metric signature  $-+++$,
\end{convention}
\begin{convention}
 units such that  $c=8\pi G =1$, where $G$ is Newton's 
gravitational constant and $c$   the speed of light in vacuum.
\end{convention}
\begin{remark}
Notice that (\ref{def_enthalpy}) assumes that $n \neq 0$. 
This is indeed the case for matter traveling at subluminal speeds. In the case of radiation,
we can take $n$ to be the particle number density instead of the rest mass density. This requires
some dimensional adjustments in the equations, which can be done 
via the introduction of a universal mass parameter that we can take to be, for example,
the proton mass\footnote{This seems a good choice
because in our present formulation the viscosity in, say,
an ultra-relativistic gas of electrons would differ from
the viscosity in a ultra-relativistic gas of muons at the
same temperature by a factor of about $200$,
just due to their mass difference, whereas if we choose a universal
mass this difference would disappear.
Since electrons and muon
interact in virtually the same way, and one would think that
viscosity has to do with interactions, it seems to be  reasonable
to introduce such an universal mass.}. With these considerations in mind, we shall assume for the entirety of the paper that $n > 0$.
\end{remark}
The stress-energy tensor (\ref{Lichnerowicz_stress}) was first introduced by Lichnerowicz
\cite{LichnerowiczBookGR}. Its motivation is given in section
\ref{section_motivation}.
 In this paper we shall study the case when
the only contribution to viscosity is given by the bulk term, so  
from now on we set $\sv = \hc = 0$, hence
\begin{gather}
T_{\al\be} = (p + \varrho) u_\al u_\be + p g_{\al\be} 
-\bv\pi_{\al\be} \nabla_\mu C^\mu,
\label{Lichnerowicz_stress_bulk}
\end{gather}
or we can write
\begin{gather}
T_{\al\be} = t_{\al\be} - \bv I_{\al\be},
\nonumber
\end{gather}
where 
\begin{gather}
t_{\al\be} = (p + \varrho) u_\al u_\be + p g_{\al\be}
\label{perfect_stress}
\end{gather}
is the stress-energy tensor for a perfect fluid,
and
\begin{gather}
I_{\al \be} = \pi_{\al\be} \nabla_\mu C^\mu
\label{imperfect_stress}
\end{gather}
is the contribution from viscosity.
We shall refer to fluid models based on (\ref{Lichnerowicz_stress}), or (\ref{Lichnerowicz_stress_bulk}) 
in the present case, as the 
\textbf{Lichnerowicz formulation of relativistic viscous fluids.}

The quantities $p$, $\varrho$, and $n$ are not all independent.
$\varrho$ and $n$ are related by 
\begin{gather}
\varrho = n(1 + \ve)
\label{density_mass_internal_energy}
\end{gather}
where $\ve$ is a scalar function representing the specific internal energy of the fluid. 
(\ref{density_mass_internal_energy}) expresses the fact the total energy density
is composed of the rest mass and the internal energy. $n$ and $\ve$ have to further satisfy
the first law of thermodynamics, which can be expressed as
\begin{gather}
d\ve = T \, ds + \frac{p}{n^2} \, dn,
\label{first_law}
\end{gather}
for some functions $s$ and $T$ that are interpreted, respectively, as the specific entropy and the
temperature of the fluid. For future reference we notice that
 (\ref{def_enthalpy}) and (\ref{density_mass_internal_energy}) give
\begin{gather}
F = 1 + \epsilon + \frac{p}{n},
\label{specific_enthalpy_2}
\end{gather}
and that  (\ref{first_law})
can be written as
\begin{gather}
dp = n\, dF - n T \, ds.
\label{first_law_2}
\end{gather}

\begin{assumption}
We shall assume for the entirety of the paper that $T > 0$. 
\end{assumption}
The above thermodynamic quantities are further linked by 
an equation of state
that relates two or more of them. Thus we assume a given relation (see remark \ref{remark_ind_thermo_var})
\begin{gather}
p = p(\varrho, F).
\label{equation_state_general}
\end{gather}
We shall also assume that $\zeta$ is given as a general function of $\varrho$ and $F$,
\begin{gather}
\zeta = \zeta(\varrho, F).
\label{zeta_function}
\end{gather}
\begin{remark}
In general only two of the thermodynamic variables can be treated as independent, with the remaining
ones determined by the first law of thermodynamics, the equation of state, and the equations of motion
(see below and 
section \ref{section_basic}). On physical grounds one assumes that the relations among these quantities
are invertible, so that the choice of which two variables are considered independent,
such as $\varrho$ and $F$ in (\ref{equation_state_general}) and (\ref{zeta_function}), it is a matter of convenience.
\label{remark_ind_thermo_var}
\end{remark}
One defines the entropy current as\footnote{We are not including a term in the heat flux because we assume
$\kappa=0$.} 
\begin{gather}
S^\mu = snu^\mu. 
\label{entropy_current_def}
\end{gather}
The second law of thermodynamics states that
\begin{gather}
\nabla_\mu S^\mu \geq 0,
\nonumber
\end{gather}
which, in light of (\ref{rest_mass_eq}), reads
\begin{gather}
u^\mu \nabla_\mu s \geq 0.
\label{second_law}
\end{gather}
We remark that (\ref{second_law}) is not a consequence of the previous equations and has
to be verified as implied by the dynamics.

The vorticity of a relativistic fluid (perfect or viscous) is defined as
(notice that our convention differs from \cite{ChristodoulouShocks,RezzollaZanottiBookRelHydro})
by a sign)
\begin{gather}
\Om_{\al\be} = \nabla_\al C_\be - \nabla_\be C_\al.
\label{vorticity}
\end{gather}
As expected, in relativity the vorticity is a  $2$-form defined in space-time rather than only in space,
as in non-relativistic mechanics. More importantly, notice that $\Om$ 
is defined in terms of the $C$ and not $u$, fundamentally contrasting with the definition 
of the vorticity in the non-relativistic setting. This is because the Kelvin circulation theorem,
believed to encode important dynamical properties of fluids, would otherwise 
not hold in the relativistic setting \cite{RezzollaZanottiBookRelHydro}. 
A fluid is called \textbf{irrotational} when $\Om = 0$.

We now turn our attention to the equations of motion.
The starting point is Einstein's equations coupled to 
(\ref{Lichnerowicz_stress_bulk}),
\begin{gather}
R_{\al\be} - \frac{1}{2} R g_{\al\be} + \La g_{\al\be} = T_{\al\be},
\nonumber
\end{gather}
where $R_{\al\be}$ and $R$ are, respectively, the Ricci and scalar curvature for the metric
$g_{\al\be}$, $\La$ is the cosmological constant, and
$T_{\al\be}$ is given by  (\ref{Lichnerowicz_stress_bulk}). 

It is convenient to rewrite
Einstein's equations as
\begin{gather}
R_{\al\be} = T_{\al \be} - \frac{1}{2}T g_{\al\be} + \La g_{\al\be},
\label{Einstein_eq}
\end{gather}
where $T = T^\al_\al$.
Einstein's equations and the Bianchi identities imply
\begin{gather}
\nabla^\al T_{\al\be} = 0.
\label{div_T_eq}
\end{gather} 
 Thus,
(\ref{div_T_eq}) is a necessary condition for (\ref{Einstein_eq}) to hold.
In situations where one neglects any gravitational dynamics caused by the presence of the fluid,
i.e., 
the so-called test fluid approximation, one still imposes  all
the previously stated relations, including (\ref{div_T_eq}), except for Einstein's 
equations themselves, i.e., (\ref{Einstein_eq}). In this case 
a metric $g$ is given {\it a priori}, and one thinks of the space-time $\st$ endowed with the metric 
$g$, $(\st,g)$, as a fixed background where the fluid dynamics takes place. While any 
$(\st,g)$ can in principle be chosen, the physically relevant scenario for the test fluid
approximation is when $(\st, g)$ satisfies the vacuum Einstein's equations.

The equations of motion consist of
 (\ref{rest_mass_eq}), (\ref{Einstein_eq}), and (\ref{div_T_eq}),
supplemented by the relations 
   (\ref{normalization_u}),  (\ref{first_law}), (\ref{equation_state_general}) and (\ref{zeta_function}). 
 We shall refer to these equations
as the \textbf{Einstein-Navier-Stokes system}.

\begin{remark}
Differentiating (\ref{normalization_u}), one obtains
\begin{gather}
u^\al \nabla_\be u_\al = 0.
\label{der_normalization_u}
\end{gather}
The reader should keep (\ref{normalization_u}), (\ref{projection_u}), and (\ref{der_normalization_u}) 
in mind as they are frequently used in the calculations throughout the paper.
\end{remark}

For a quick discussion of thermodynamic properties of relativistic fluids, 
see, e.g., \cite{DisconziRemarksEinsteinEuler} or \cite{FriRenCauchy}. A very thorough
and up-to-date
account of relativistic fluids, which includes a review of viscosity in relativity,
can be found in the monograph \cite{RezzollaZanottiBookRelHydro}.
Mathematical studies of Lichnerowicz's formulation of the Einstein-Navier-Stokes system 
with shear viscosity (i.e., using (\ref{Lichnerowicz_stress})) have been
carried out in \cite{DisconziViscousFluidsNonlinearity, DisconziCzubakNonzero,
PichonViscous}. The mathematical properties of relativistic perfect fluids are discussed in
the monographs
\cite{AnileBook, ChoquetBruhatGRBook, ChristodoulouShocks, Lichnerowicz_MHD_book}.
See also the related papers
\cite{AnilePennisiMHD,Choquet-BruhatFluidsExistence,Lich_MHD_paper,Lich_fluid_1, Lich_fluid_2}.

Some of the primary testable physical predictions of models for relativistic viscosity lie in the field of cosmology.
In general, the effect of bulk viscosity on a fluid is to decrease its effective pressure, and a fluid with
a sufficiently negative pressure can serve as a source for the observed acceleration of the
expansion of the universe.
This possibility has been extensively explored in connection with earlier models of viscosity in Refs.
\cite{Avelino,Balakin,Gagnon,Hipolito1,Hipolito2,Colistete,Velten,Zimdahl}.  Preliminary results for
the cosmological effects of Lichnerowicz viscosity are presented in Ref. \cite{Disconzi_Kephart_Scherrer_2015},
and generalized in \cite{MontaniLichnerowiczViscosityBianchi}.

We finish this introduction with a few general remarks regarding some of the goals of this work.
As already mentioned, there are different proposals for describing relativistic viscous fluids in the literature, and
this is not a settled issue.  We do not know
which of the several formulations is the correct one (in contrast to the confidence with which we can say
that the Einstein-Euler system is the correct formulation for relativistic perfect fluids).
Hence, we believe that it is important to investigate Lichnerowicz's proposal, especially because it is 
less known in the community.
Now, precisely because Lichnerowicz's formulation is less known, it has been less developed compared
to other theories. This of course does not mean the Lichnerowicz's theory is less interesting or less deserving of attention
(at least given what we know about it so far), but it does mean that many basic questions remain open.

Given the above, one of the main goals of the paper is to bring attention to Lichnerowicz's theory, and show how one
can derive some interesting properties of fluids and cosmological models from it. Therefore, we have not strived for
complete generality
and will make several simplifying assumptions below.

Some of the topics here discussed may also be of interest beyond the general relativity and cosmology communities,
such as researchers working on heavy-ion collisions or mathematicians working on mathematical fluids dynamics. Therefore,
we attempted to make our presentation as self-contained as possible, at the expense of sometimes repeating ideas
known to specialists.

\section{Summary of results} Here we briefly summarize the results of the paper. The statements
below are intended solely as a quick outline of this work and therefore, do not mention
 technical assumptions or subtleties. More precise statements are found in the sections below.

Regarding the formulation of relativistic viscous fluids 
derived from (\ref{Lichnerowicz_stress_bulk}):
\begin{itemize}
\item It  does not fall into the general class of first order theories proved to be acausal and unstable by 
Hiscock and Lindblom \cite{Hiscock_Lindblom_instability_1985}. See section \ref{section_theories}.
We need to be more specific by what we mean here. First, (\ref{Lichnerowicz_stress}) 
does not satisfy the assumptions of the stress-energy tensor employed in  \cite{Hiscock_Lindblom_instability_1985}.
While this prevents the results of \cite{Hiscock_Lindblom_instability_1985} from being invoked, it of course
does not in itself mean that acausality and instabilities are absent, although it opens that possibility. But a bit more can be said.
If one repeats the linearization procedure of \cite{Hiscock_Lindblom_instability_1985} applied to (\ref{Lichnerowicz_stress}) ,
one arrives at a system for which modes cannot be shown to grow using the same techniques of \cite{Hiscock_Lindblom_instability_1985}.
Again, this does not indicate that exponentially growing modes are absent, but it does suggest a significantly different dynamics that cannot
be shown to be acausal in a more or less straightforward manner. A full analysis of perturbations of 
Lichnerowicz's formulation near equilibrium will be the subject of a forthcoming work \cite{BemficaDisconziNoronha-InPrep}.
\item The equilibrium states for (\ref{Lichnerowicz_stress_bulk}) include those of Eckart and MIS theories.
See section \ref{section_equilibrium}.
\item There are natural sufficient conditions that lead to non-negative entropy production.
See claim \ref{claim_entropy_production}.
\item It yields the correct non-relativistic limit. See section \ref{section_nr_limit}.
\item The system of Einstein's equations coupled to (\ref{Lichnerowicz_stress_bulk}) is causal if
the fluid is irrotational and other mild conditions are satisfied. See claim \ref{claim_causality}.
\end{itemize}
Regarding applications of (\ref{Lichnerowicz_stress_bulk}) to cosmology:
\begin{itemize}
\item For a generic dark fluid, the presence of viscosity leads to a decrease in
the effective equation of state parameter, $w_{eff}$, relative to its
value in the absence of viscosity, $w$.
\item Whether the effect of viscosity is dominant or subdominant at late times
depends on the value of $w$ and the scaling of the viscosity parameter
with the density of the fluid.
\item For a wide choice of parameter values, Lichnerowicz viscosity drives the
dark fluid toward phantom-like behavior ($w_{eff} < -1$) and a future
big-rip singularity.
\end{itemize}

\section{Similarities and differences with other theories of relativistic viscosity\label{section_theories}}

In this section we discuss how Lichnerowicz's approach based
on (\ref{Lichnerowicz_stress}), and our work in particular, fits within the broader 
context of viscosity in relativity. 

The following stress-energy tensor for relativistic viscous fluids was introduced by
Eckart in 1940 \cite{EckartViscous}:
\begin{align}
\begin{split}
T^E_{\al\be}&  = (p + \varrho) u_\al u_\be + p g_{\al\be} 
-(\bv - \frac{2}{3} \sv) \pi_{\al\be} \nabla_\mu u^\mu \\
& - \sv \pi_\al^\mu \pi_\be^\nu
(\nabla_\mu u_\nu + \nabla_\nu u_\mu) - \hc (q_\al u_\be + q_\be u_\al).
\end{split}
\label{Eckart_stress}
\end{align}
All quantities in (\ref{Eckart_stress}) are as defined in section \ref{section_intro}.
Approaches based on (\ref{Eckart_stress}) are called Eckart theories.
We remark that (\ref{Eckart_stress}) is commonly written as
\begin{gather}
T^E_{\al\be} = (p + \varrho) u_\al u_\be + p g_{\al\be} 
-\bv  \pi_{\al\be} \nabla_\mu u^\mu - 2\sv \si_{\al \be}  - \hc (q_\al u_\be + q_\be u_\al),
\nonumber
\end{gather}
where
\begin{gather}
\si_{\al\be} = \nabla_{(\al} u_{\be)} + a_{(\al} u_{\be)} - \frac{1}{3} \pi_{\al\be} \nabla_\mu u^\mu,
\nonumber
\end{gather} 
is called the shear-tensor, with $a_\al = u^\mu \nabla_\mu u_\al$ being the acceleration.

Eckart motivated (\ref{Eckart_stress}) from basic assumptions of relativity and thermodynamics.
Although such assumptions are not sufficient to uniquely determine the stress-energy tensor, the form
(\ref{Eckart_stress}) is, to quote Eckart, ``strongly indicated, if not uniquely determined."

Hiscock and Lindblom have showed that the equations of motion derived from (\ref{Eckart_stress})
suffer from several pathologies, including acausality and instabilities that are incompatible with 
observation \cite{Hiscock_Lindblom_instability_1985, Hiscock_Lindblom_pathologies_1988}.
Even with these inconsistencies, the Eckart theory was nonetheless used, fully or partially,  
with different degrees of success, in the study of
 neutron stars,
supernovae, and in some models of viscous cosmology. 
Without being exhaustive, 
we provide the following list of references to this important topic:
\cite{
Avelino,
Balakin,
Barrow1,
Barrow2,
Barrow,
Barrowbook,
BrevikBigRip,
BG,
Cooketal2003,
Disconzi_Kephart_Scherrer_2015,
Dosetal,
DosTsa,
Duez_review,
DuezetallEinsteinNavierStokes,
Gagnon,
Geroch-RelativisticDissipative, 
GerochHyperbolicTheoriesNotViable, 
GerochLindblomDivergenceType,
GerochLindblomCausal,
Gron,
HZS,
HS,
HerreraPavon-HyperbolicTheoriesDissipation,
Herr_axially,
Herr_axially_shear,
Hipolito1,
Hipolito2,
Colistete,
Kreiss_et_al,
LiBarrow,
LiuMullerRuggeri-RelThermoGases,
Lovelace_Duez_et_al,
MaartensDissipative,
MontaniLichnerowiczViscosityBianchi,
Murphy,
Nagy_et_all-Hyperbolic_parabolic_limit,
Pad,
Pahwa_FLRW,
Pal_et_al,
Pal_Reula_Rezzolla,
Pavon-CaseHyperbolicDissipation,
Piattella_et_al,
Reula_et_al-CausalStatistical,
RezzollaZanottiBookRelHydro,
Saijo,
SorBran,
TreciokasEllisViscosity,
VeltenSchwarz,
Velten,
Waga,
WeinbergViscosityCosmology,
Weinberg_GR_book,
Zimdahl}.

The  Mueller-Israel-Stewart (MIS) theory
\cite{MIS-2, MIS-3, MIS-5, MIS-6, MIS-1, MIS-4}
is probably the most well-studied attempt to overcome
the lack of causality of Eckart's theory. It consists of 
a systematic application of the ideas of relativistic extended irreversible thermodynamics
\cite{JouetallBook, MuellerRuggeriBook}. In these theories, the viscous contributions
to the stress-energy tensor are assumed to be new variables in the theory rather than given in
terms of $u$ and the other thermodynamic variables. Equations of motion for the new variables
are then chosen so that the second law of thermodynamics is satisfied. We briefly review
the basic constructions, following mostly the discussion in 
\cite{RezzollaZanottiBookRelHydro}.

Consider a stress-energy tensor of the form
\begin{gather}
\widetilde{T}_{\al\be} = t_{\al\be} + \pi_{\al\be} \Pi + \Pi_{\al\be} + Q_\al u_\be + Q_\be u_\al.
\nonumber
\end{gather}
The terms $\Pi$, $\Pi_{\al\be}$, and $Q_\al$ correspond to the dissipative contributions
to the stress-energy tensor. If we set
\begin{gather}
\Pi = -\bv \nabla_\mu u^\mu,
\nonumber
\end{gather}
\begin{gather}
\Pi_{\al\be} = - \sv \pi_\al^\mu \pi_\be^\nu
(\nabla_\mu u_\nu + \nabla_\nu u_\mu - \frac{2}{3} \nabla_\mu u^\mu) 
\nonumber
\end{gather}
and 
\begin{gather}
Q_\al = -\hc q_\al,
\nonumber
\end{gather}
we recover Eckart's stress-energy tensor (\ref{Eckart_stress}). However, instead of 
adopting these
definitions, the quantities $\Pi$, $\Pi_{\al\be}$, and $Q_\al$
are treated as new variables in the MIS theory. Since new independent variables are being introduced,
extra equations of motion also have to be introduced in order to close the system of equations it generates. 
The new equations are postulated based on the following argument.

In Eckart's theory, when $\hc \neq 0$, the entropy current (\ref{entropy_current_def}) becomes
\begin{gather}
S^\al = snu^\al + \hc \frac{q^\al}{T}.
\label{entropy_current_first_order}
\end{gather}
In the MIS theory, one postulates an entropy current of the form
\begin{align}
\begin{split}
S^\al = snu^\al +  \frac{Q^\al}{T}
- (\be_0 \Pi^2 + \be_1 Q_\mu Q^\mu + \be_2 \Pi_{\mu\nu}\Pi^{\mu\nu} ) \frac{u^\al}{2T}
+ \al_0 \frac{\Pi Q^\al}{T} + \al_1 \frac{\Pi^{\al\mu} Q_\mu}{T},
\end{split}
\label{entropy_current_second_order}
\end{align}
for some coefficients $\be_0$, $\be_1$, $\be_2$, $\al_0$, and $\al_1$.
Next, we compute $\nabla_\mu S^\mu$ and seek the simplest relation, linear
in the variables $\Pi$, $\Pi_{\al\be}$, and $Q_\al$, which assures that the
second law of thermodynamics, (\ref{second_law}), is satisfied. In this way we
obtain (see, e.g., \cite{MaartensLecturesDissipative, RezzollaZanottiBookRelHydro})
\begin{align}
\begin{split}
\tau_0 u^\mu \nabla_\mu \Pi + \Pi = -\bv \nabla_\mu u^\mu -\frac{1}{2} \bv
T \nabla_\mu ( \frac{\tau_0}{\bv T} u^\mu ) \Pi,
\end{split}
\label{eq_motion_Pi}
\end{align}
\begin{align}
\begin{split}
\tau_1 \pi^\la_\mu u^\nu \nabla_\nu Q_\la + Q_\mu = -\hc T(\pi^\nu_\mu \nabla_\nu \ln T + 
a_\mu) -\frac{1}{2} \hc T^2 \nabla_\nu ( \frac{\tau_1}{\hc T^2} u^\nu ) Q_\mu,
\end{split}
\label{eq_motion_Q}
\end{align}
and 
\begin{align}
\begin{split}
\tau_2 \pi^\la_\mu \pi^\si_\nu u^\al \nabla_\al \Pi_{\la\mu} = -2\sv \si_{\mu\nu}
 -\frac{1}{2} \sv  T \nabla_\al ( \frac{\tau_2}{\sv T} u^\al ) \Pi_{\mu\nu},
\end{split}
\label{eq_motion_Pi_tensor}
\end{align}
where the non-negative coefficients of bulk and shear viscosity and heat conduction,
$\bv$, $\sv$, and $\hc$, respectively, have been introduced and 
\begin{gather}
\tau_0 = \bv \be_0, \, \tau_1 = \hc T \be_1, \, \text{ and } \, \tau_2 = 
2\sv \be_2,
\nonumber
\end{gather}
are characteristic relaxation times.

Approaches to relativistic viscosity such as Eckart's or Lichnerowicz's, where
the dynamic content of the theory is encoded in the stress-energy tensor and
an entropy current of the form (\ref{entropy_current_first_order}) is adopted, 
are known as first order theories, indicating that (\ref{entropy_current_first_order})
incorporates only first order deviations from thermodynamic equilibrium. Approaches like the MIS theory
where further equations of motion are postulated for the dissipative variables, now treated
as independent, such as (\ref{eq_motion_Pi}), (\ref{eq_motion_Q}), and (\ref{eq_motion_Pi_tensor}),
and where the entropy current is given by (\ref{entropy_current_second_order}), are known
as second order theories, referring to the fact that (\ref{entropy_current_second_order})
incorporates second order deviations from thermodynamic equilibrium.

The linearization about equilibrium
of the MIS theory has been shown to be causal \cite{Hiscock_Lindblom_stability_1983}.
However, when the non-linear theory is considered, the MIS can also lead to a lack of 
causality \cite{Hiscock_Lindblom_pathologies_1988}. 
To be fair, in \cite{Hiscock_Lindblom_pathologies_1988} such a loss of causality
happens under extreme physical conditions unlikely to be met by most physically realistic systems.
On the other hand, the system studied in \cite{Hiscock_Lindblom_pathologies_1988}
is significantly simple, namely,
 only heat conduction is present, so that the bulk and shear viscosities are zero,
and it assumes planar symmetry. In particular, it is not clear
under which conditions equations
(\ref{eq_motion_Pi}), (\ref{eq_motion_Q}), and (\ref{eq_motion_Pi_tensor})
remain causal when $\sv$ and $\bv$ are not zero. This is a particularly delicate
question in view of the projection operators $\pi^\la_\mu \pi^\si_\nu$ in
(\ref{eq_motion_Pi_tensor}) which typically lead to a failure of hyperbolicity or, in slightly
better situations, to weakly hyperbolic operators (see, e.g., 
\cite{DisconziCzubakNonzero,DisconziViscousFluidsNonlinearity} and references therein for more discussion
on this technical point; see also section \ref{section_naive}). Further discussion
about conditions that give causality or non-causality of the MIS theory can be found in 
\cite{RezzollaZanottiBookRelHydro}. In this regard,
we notice that  Rezzolla and Zanotti \cite{RezzollaZanottiBookRelHydro} conclude their detailed discussion of relativistic viscous fluids 
pointing out that ``the construction of a formulation that is cast in a divergence-type\footnote{Which is
a refinement of the extended irreversible thermodynamics approach on which the MIS theory
is based.} is not, per se, sufficient to guarantee hyperbolicity" (and hence causality). 
In passing, we notice that although the MIS theory is commonly referred to
 as a ``causal dissipative relativistic" theory, this is perhaps a misnomer.

Thus, while it is fair to say that second order theories in general, and the MIS theory in particular,
have been an important step forward,  their 
causality properties are not yet fully understood.  Furthermore,
these theories present what is, in our opinion, undesirable features, namely:

(i) Equations (\ref{eq_motion_Pi}), (\ref{eq_motion_Q}), and (\ref{eq_motion_Pi_tensor}) are ultimately
arbitrary. As said, they are the simplest equations one can impose in order to guarantee (\ref{second_law}),
but many other relations (for instance, non-linear in the dissipative variables) that ensure
(\ref{second_law}) can  equally well be postulated (see also the discussion in \cite{Disconzi_Kephart_Scherrer_2015}),
and none of these choices seem to be suggested by the structure of general relativity. This is in
sharp contrast with the standard framework of general relativity, where the dynamical content is usually
fully encoded in Einstein's equations. Even when further equations have to be postulated in general relativity,
as in the case of (\ref{rest_mass_eq}), they traditionally have 
an intrinsic physical interpretation that is intended to provide some a posteriori 
justification of a logically independent requirement.

(ii) The physical content of the $\al_i$ and $\be_i$ coefficients 
in (\ref{entropy_current_second_order}) is not apparent (see, however, the 
discussion in \cite{RezzollaZanottiBookRelHydro}).

(iii) Since the dissipative quantities in second order theories are now independent variables,
it is not at all clear that the classical Navier-Stokes equations can be obtained as a non-relativitistic
limit of second order theories without further, perhaps unnatural, assumptions (compare with section
\ref{section_nr_limit}).
There have been works showing how solutions to theories of relativistic viscosity are near their non-relativistic
limits (see, e.g., \cite{Geroch-RelativisticDissipative, Kreiss_et_al, Nagy_et_all-Hyperbolic_parabolic_limit} and references therein).
By the non-relativistic limit, however, we mean a stronger condition, namely, that 
the equations themselves reduce to the classical equations in the limit, since
this is indeed what happens for the relativistic Euler equations and most matter models of interest.

(iv) To the best of our knowledge, the well-posedness and causality of the MIS theory
and other second order formulations \emph{coupled} to Einstein's equation is 
not available in the literature, with all the aforementioned
results restricted to the case of a background metric (in fact, the Minkowski metric in most cases). 
It is perhaps worth noticing that simply showing that a given second order theory admits, in certain situations,
a representation as a first order symmetric hyperbolic system (as it is sometimes the case for second
order theories) is not enough to guarantee
well-posedness and causality when one couples to Einstein's equations,
 although that will be the case when the coupling happens at lower order. In any case, requiring
the equations to always form a symmetric hyperbolic system would be
too stringent a requirement to impose on physical theories.

(v) The MIS theory does not admit strong shock wave solutions. See 
\cite{HiscockShocksMIS} for details (see also the discussion in \cite{GerochLindblomCausal}).

These considerations make the exploration of alternative
approaches worthwhile. The equations studied by the authors point 
toward the mitigation of several of these difficulties. 

Although we have not yet treated the most general case,
the Lichnerowicz formulation of relativistic viscous fluids has been
shown to be well-posed and causal under the following circumstances:

(a) For irrotational fluids satisfying the ``incompressibility" condition $\nabla_\mu C^\mu = 0$
with no heat conduction or bulk viscosity 
 \cite{DisconziViscousFluidsNonlinearity}, but
with shear viscosity;

(b) For fluids with vorticity and shear viscosity, but with no bulk viscosity or heat conduction, and
 under restrictions on the initial data\footnote{Although the restrictions on the initial data
in \cite{DisconziCzubakNonzero} are rather strong, it will be shown in a forthcoming work how these
restrictions can be relaxed.}
\cite{DisconziCzubakNonzero};
 
(c) For irrotational fluids with bulk viscosity but no shear
viscosity or heat conduction, and very mild assumptions on the initial data
(see section \ref{section_ENS_causal}).

It should be noticed that in all the three cases (a), (b), and (c), 
the full non-linear problem has been addressed,
\emph{including coupling to Einstein's equations,} without ``near equilibrium" assumptions.
In particular, the results (a) and (b) do deal with the ``dangerous" projection terms
$\pi_\al^\mu \pi_\be^\nu$ in $T_{\al\be}$ that tend  to spoil causality,
as mentioned above.  It should be pointed out that causality of Lichnerowicz's formulation 
in a fixed background, under the same assumptions as in (a), (b), or (c), can be shown by exactly the same
techniques used in  \cite{DisconziCzubakNonzero,DisconziViscousFluidsNonlinearity}. The analytic
Cauchy problem for (\ref{Lichnerowicz_stress}) has been studied in \cite{PichonViscous}.

It should also be pointed out that Lichnerowicz's stress-energy tensor
(\ref{Lichnerowicz_stress}) does not satisfy the assumptions
of the instability results proved by Hiscock and Lindblom
 \cite{Hiscock_Lindblom_instability_1985}, which apply to a large 
class of first order theories (including Eckart's theory).
Thus, the simple modification
proposed by Lichnerowicz, namely, replacing the velocity by the dynamic velocity in the viscous
part of the stress-energy tensor, has far-reaching consequences. In particular, one
can circumvent Hiscock and Lindblom's result upon adopting Lichnerowicz's proposal, which is what we do here.

The above remarks are not intended to claim that Lichnerowicz's proposal  
is better than the more studied MIS approach (or better than other second order theories), 
but rather to highlight how little is known about 
viscosity in general relativity and to point out that studying possible alternatives
to the MIS  formulation is a pressing issue. Despite the fact that Lichnerowicz first wrote (1.1) about sixty years ago
\cite{LichnerowiczBookGR}, with the exception of Pichon's work \cite{PichonViscous}, which did 
not address questions of causality and hyperbolicity, it has not been until 
very recently that Einstein's equations coupled to (\ref{Lichnerowicz_stress}) have been investigated 
\cite{DisconziCzubakNonzero, DisconziViscousFluidsNonlinearity}.

One curious feature of Lichnerowicz's formulation is that for a fluid
with equation of state $p = -\varrho$, the viscous contributions vanish. This is
of interest in the cosmological applications discussed in part \ref{part_applications}.

Recently Freist\"uhler and Temple have proposed yet another stress-energy tensor
for relativistic viscous fluids that leads to a causal dynamics for certain values of 
the viscosity coefficients and an equation of state for pure radiation \cite{TempleViscous}.
Their results also include a description of strong shock waves, in contrast to (v) listed above.
The general point of view adopted in \cite{TempleViscous} is one we share. 
As we saw, the equations of the MIS theory are obtained upon imposing that entropy production be
non-negative, for all values of the thermodynamic variables.
In \cite{TempleViscous}, the authors point out that this is too stringent a requirement, remarking that 
``dissipative fluid dynamics is a perturbative theory, intended for small
dissipation, the case when gradients are close to those that occur in the inviscid limit. That is,
entropy production need not necessarily be positive on gradients far from this case" \cite{TempleViscous}.
Thus, they seek a stress-energy tensor that produces a causal dynamics, proving 
a posteriori that the second law thermodynamics holds. Here, 
as in \cite{DisconziCzubakNonzero,DisconziViscousFluidsNonlinearity, Disconzi_Kephart_Scherrer_2015}
a similar philosophy is adopted, namely, neither causality nor the second law are guaranteed {\it a priori}. Instead,
one tries to find reasonable conditions under which such properties hold, even if these conditions
will not describe the complete state space of the theory. This may be a more laborious task, since
one condition that was automatically verified in the MIS theory now has to be derived from
the dynamics. On the other hand, it gives us more flexibility in trying to balance adequate conditions that
lead to causality with those necessary for non-negative entropy production. Another first-order viscous theory
that deserves mention is that of V\'an and Bir\'o \cite{VanFirstOrder}.

Even a brief discussion of the literature as presented here would be incomplete without mentioning
the work of Geroch and Lindblom \cite{GerochLindblomCausal}. There, a fluid theory is defined
in rather general terms, and it is shown that the MIS, Eckart, Landau \cite{LandauLifshitzFluids}, and some of the so-called
divergence-type theories  \cite{GerochLindblomDivergenceType, LiuMullerRuggeri-RelThermoGases}, all fit in their 
formalism\footnote{Landau's theory is a theory resembling Eckart's; a discussion of divergence-type theories can be found in 
the previous references and also in \cite{MuellerRuggeriBook, RezzollaZanottiBookRelHydro}. These theories will
not be discussed in this paper.}. 

Geroch and Lindblom provide sufficient conditions guaranteeing that a fluid theory
is well-posed and causal. The task then becomes to determine when a given theory that fits their general formalism
also satisfies those additional conditions. This can only be done in a case-by-case analysis. While the MIS is a fluid theory
as defined in \cite{GerochLindblomCausal}, it does not satisfy the additional conditions ensuring causality (although, 
as already mentioned, it can be verified that the MIS is causal under certain circumstances by a direct analysis 
of its equations of motion). Other topic addressed in \cite{GerochLindblomCausal} are the stability of solutions
and shock waves.

\part{THE FLUID MODEL}

\section{Basic features of the fluid model \label{section_basic}}

In this section we discuss some basic features of the fluid model under study.

\subsection{Equations of motion \label{section_eq_motion}}
We can decompose $\nabla^\al T_{\al\be} = 0$ in the direction parallel and orthogonal to 
$u$, i.e.,  $u^\be \nabla^\al T_{\al\be} = 0$ and 
$\pi^{\be\ga} \nabla^\al T_{\al\ga} = 0$. The former gives 
\begin{gather}
u^\mu \nabla_\mu \varrho + (p+\varrho) \nabla_\mu u^\mu 
-\zeta [ F (\nabla_\mu u^\mu)^2 + \nabla_\mu u^\mu u^\al \nabla_\al F] = 0.
\label{rho_equation}
\end{gather}
In order to obtain $\pi^{\be\ga} \nabla^\al T_{\al\ga} = 0$ in a form suitable for our purposes, 
we use 
(\ref{vorticity})
and  the standard identity
\begin{gather}
\nabla_\al \nabla_\be v^\ga - \nabla_\be \nabla_\al v^\ga =
R_{\al\be\mss \la}^{\mss\mss\ga} v^\la
\nonumber
\end{gather}
 to obtain
\begin{align}
\begin{split}
\nabla^\al \Om_{\al\be} & = \nabla^\al \nabla_\al C_\be - \nabla^\al \nabla_\be C_\al \\
& = \nabla^\al \nabla_\al C_\be - \nabla_\be \nabla_\al C^\al - R_{\be \al} C^\al,
\nonumber
\end{split}
\end{align}
or upon rearranging 
\begin{align}
\begin{split}
\nabla_\be \nabla_\al C^\al  & = \nabla^\al \nabla_\al C_\be - \nabla^\al \Om_{\al\be} - R_{\be \al} C^\al.
\nonumber
\end{split}
\end{align}
But
\begin{gather}
\nabla^\al \nabla_\al C_\be = u_\be \nabla^\al \nabla_\al F 
+ 2 \nabla_\al F \nabla^\al u_\be + F \nabla^\al \nabla_\al u_\be,
\nonumber
\end{gather}
so 
\begin{align}
\begin{split}
\pi^{\be\ga} \nabla^\al I_{\al\ga} 
& = F \nabla^\mu \nabla_\mu u^\be - F u^\be \nabla_\mu u_\al \nabla^\mu u^\al 
+ F \nabla_\mu u^\mu u^\al \nabla_\al u^\be + u^\mu \nabla_\mu F u^\al \nabla_\al u^\be 
\\
& + 2 \nabla_\mu F \nabla^\mu u^\be - \pi^{\be\al}(\nabla^\mu \Om_{\mu \al} + F R_{\al \mu}u^\mu).
\end{split}
\nonumber
\end{align}
From this and $\pi^{\be\ga} \nabla^\al T_{\al\ga} = 0$ we deduce the following 
equation
\begin{align}
\begin{split}
& \zeta F \nabla^\mu \nabla_\mu u^\be 
+ \zeta F \nabla_\mu u^\mu u^\al \nabla_\al u^\be 
- \zeta F u^\be \nabla_\mu u_\al \nabla^\mu u^\al 
+ \zeta u^\mu \nabla_\mu F u^\al \nabla_\al u^\be 
\\
& + 2 \zeta \nabla_\mu F \nabla^\mu u^\be - \zeta \pi^{\be\al}(\nabla^\mu \Om_{\mu \al} + F R_{\al \mu}u^\mu)
+ (F \nabla_\mu u^\mu + u^\mu \nabla_\mu F) \pi^{\be \al} \nabla_\al \zeta \\
& - (p + \varrho) u^\al \nabla_\al u^\be - \pi^{\be \al} \nabla_\al p = 0.
\end{split}
\label{u_equation}
\end{align}
Our system describing a relativistic viscous fluid consists
of equations  (\ref{rest_mass_eq}), (\ref{Einstein_eq}), (\ref{rho_equation}), and (\ref{u_equation}),
supplemented by the relations 
  (\ref{normalization_u}),  (\ref{first_law}),  (\ref{equation_state_general}) and (\ref{zeta_function}). 
 One recalls that $\Om$ is given
 in terms of $u$ and $F$ 
 by (\ref{def_C}) and  (\ref{vorticity}). In accordance to what was discussed in the introduction,
 we shall refer to these equations
as the \textbf{Einstein-Navier-Stokes system}.

\subsection{Equations of motion for $p = w \varrho$}

In this section we assume an equation of state of the form
\begin{gather}
p = w \varrho, \, w \, \text{ constant},
\label{equation_state}
\end{gather}
(which will be the main focus of applications in part \ref{part_applications}), and re-write the Einstein-Navier-Stokes
system in a form suitable for our purposes in the ensuing sections.  Eq. (\ref{equation_state})
applies, for instance, to radiation ($w=1/3$) and dust ($w=0$) and is also the starting point
for many discussions of dark energy. Besides its importance in cosmology,
we focus on equation (\ref{equation_state}) due to its simplicity (recall, as mentioned in the introduction,
that we are not aiming at complete generality in this work. Results treating a general equation of state can be found
in \cite{DisconziCzubakNonzero,DisconziViscousFluidsNonlinearity}).

With (\ref{equation_state}), (\ref{rho_equation}) becomes

\begin{gather}
u^\mu \nabla_\mu \varrho + (1+w)\varrho \nabla_\mu u^\mu 
-\zeta [ F (\nabla_\mu u^\mu)^2 + \nabla_\mu u^\mu u^\al \nabla_\al F ] = 0.
\label{rho_equation_w_prelim}
\end{gather}
Using (\ref{def_enthalpy}) and (\ref{equation_state}), (\ref{rest_mass_eq}) can be written as 
\begin{gather}
\frac{\varrho}{F} u^\mu \nabla_\mu F - u^\mu \nabla_\mu \varrho - \varrho \nabla_\mu u^\mu = 0.
\label{F_equation_w_prelim}
\end{gather}
Adding (\ref{rho_equation_w_prelim}) and (\ref{F_equation_w_prelim}) produces
\begin{gather}
(1 - \zeta \frac{F}{\varrho} \nabla_\mu u^\mu) u^\al \nabla_\al F + w F \nabla_\mu u^\mu 
- \zeta \frac{F^2}{\varrho} (\nabla_\mu u^\mu )^2 = 0.
\label{F_equation_w}
\end{gather}
Using (\ref{F_equation_w_prelim}) to eliminate $u^\mu \nabla_\mu F$ from (\ref{rho_equation_w_prelim})
gives
\begin{gather}
(1 - \zeta \frac{F}{\varrho} \nabla_\mu u^\mu) u^\al \nabla_\al \varrho + (1+w) \varrho \nabla_\mu u^\mu 
-2 F \zeta (\nabla_\mu u^\mu)^2 = 0.
\label{rho_equation_w}
\end{gather}
Finally, under assumption (\ref{equation_state}), (\ref{u_equation}) reads
\begin{align}
\begin{split}
& \zeta F \nabla^\mu \nabla_\mu u^\be 
+ \zeta F \nabla_\mu u^\mu u^\al \nabla_\al u^\be 
- \zeta F u^\be \nabla_\mu u_\al \nabla^\mu u^\al 
+ \zeta u^\mu \nabla_\mu F u^\al \nabla_\al u^\be 
\\
& + 2 \zeta \nabla_\mu F \nabla^\mu u^\be - \zeta \pi^{\be\al}(\nabla^\mu \Om_{\mu \al} + F R_{\al \mu}u^\mu)
+ (F \nabla_\mu u^\mu + u^\mu \nabla_\mu F) \pi^{\be \al} \nabla_\al \zeta \\
& - (1+w) \varrho u^\al \nabla_\al u^\be - w \pi^{\be \al} \nabla_\al \varrho = 0.
\end{split}
\label{u_equation_w}
\end{align}
Therefore, when (\ref{equation_state}) holds, we can alternatively take  (\ref{F_equation_w}), (\ref{rho_equation_w}), (\ref{u_equation_w}), and (\ref{Einstein_eq}) as the Einstein-Navier-Stokes system.

\subsection{Equilibrium states\label{section_equilibrium}}

Following \cite{Hiscock_Lindblom_instability_1985,RezzollaZanottiBookRelHydro},
we define a fluid to be in \textbf{thermodynamic equilibrium} when it is stationary with respect to variations
in the thermodynamic flux given by the viscous bulk term. Mathematically this means that
\begin{gather}
I_{\al\be} = 0.
\label{equilibrium_fluid_general}
\end{gather}
Naturally, (\ref{equilibrium_fluid_general}) is automatically satisfied for perfect fluids, and our interest is
in determining thermodynamic equilibrium states when viscosity is present, so here we assume
$\zeta \neq 0$, in which case (\ref{equilibrium_fluid_general}) reads
\begin{gather}
\nabla_\mu C^\mu = 0.
\label{equilibrium_fluid}
\end{gather}
 We also notice that 
thermodynamic equilibrium 
implies no entropy production, i.e., 
\begin{gather}
\nabla_\mu S^\mu = 0,
\nonumber
\end{gather}
or, in view of (\ref{rest_mass_eq}) and $n>0$, 
\begin{gather}
u^\mu \nabla_\mu s = 0.
\label{equilibrium_s}
\end{gather}
A fluid satisfying (\ref{equilibrium_s}) is called adiabatic.

Using
\begin{gather}
u^\al \nabla_\al \varrho = (1+\epsilon) u^\al \nabla_\al n + n u^\al \nabla_\al \epsilon
\nonumber
\end{gather}
that follows from (\ref{density_mass_internal_energy}), and
\begin{gather}
(p + \varrho) \nabla_\al u^\al = -\frac{p_+\varrho}{n} u^\al \nabla_\al n,
\nonumber
\end{gather}
that follows from (\ref{rest_mass_eq}), in (recalling also (\ref{perfect_stress}))
\begin{gather}
u^\be \nabla_\al t^\al_\be = - u^\al \nabla_\al \varrho - (p+\varrho) \nabla_\al u^\al,
\label{rho_eq_perfect}
\end{gather}
produces
\begin{gather}
u^\be \nabla_\al t^\al_\be = -nu^\al \nabla_\al \epsilon + \frac{p}{n} u^\alpha \nabla_\al n,
\nonumber
\end{gather}
where we have also used (\ref{def_enthalpy}).
 Upon contracting (\ref{first_law}) with $u$, this becomes
\begin{gather}
u^\be \nabla_\al t^\al_\be = - T n u^\al \nabla_\al s.
\label{u_div_T_perfect}
\end{gather}
Thus, (\ref{rho_equation}) can be rewritten as 
\begin{gather}
-Tn u^\mu \nabla_\mu s + \zeta F (\nabla_\mu u^\mu)^2+ \zeta \nabla_\mu u^\mu u^\al \nabla_\al F = 0 .
\label{s_equation}
\end{gather}
In particular, we see that (\ref{equilibrium_s}) is automatically satisfied for a perfect fluid.

When only bulk viscosity is present, the equilibrium states of both Eckart's and the MIS theories
are characterized by
\begin{gather}
\nabla_\mu u^\mu = 0.
\label{div_u_0}
\end{gather}
Let us show (\ref{div_u_0}) also implies thermodynamic equilibrium in the case of Lichnerowicz's
formulation. If (\ref{div_u_0}) holds, then 
(\ref{rho_eq_perfect}), (\ref{u_div_T_perfect}), (\ref{s_equation}), and (\ref{rest_mass_eq}) imply
\begin{gather}
u^\mu \nabla_\mu s = 0, \, u^\mu \nabla_\mu \varrho = 0, \, \text{ and } \, u^\mu \nabla_\mu n = 0.
\nonumber
\end{gather}
Writing $p = p(\varrho, s)$ (recall remark \ref{remark_ind_thermo_var}), the above then implies
\begin{gather}
u^\mu \nabla_\mu p = \frac{\partial p}{\partial \varrho} u^\mu \nabla_\mu \varrho 
+ \frac{\partial p }{\partial s} u^\mu \nabla_\mu s = 0.
\nonumber
\end{gather}
Therefore, from (\ref{def_enthalpy})
\begin{gather}
u^\mu \nabla_\mu F = 0.
\nonumber
\end{gather}
This last equality and (\ref{div_u_0}) give (\ref{equilibrium_fluid}).

We now ask if the converse is true, i.e., if (\ref{equilibrium_fluid}) implies (\ref{div_u_0}), thus assume 
(\ref{equilibrium_fluid}). Combined with (\ref{rest_mass_eq})  this implies
\begin{gather}
u^\mu \nabla_\mu \left( \frac{n}{F} \right) = 0,
\nonumber
\end{gather}
provided that $F \neq 0$. We can assume that the equation of state is such that
$n = n(F,s)$ (see remark \ref{remark_ind_thermo_var}), so that
\begin{gather}
( F \frac{\partial n}{\partial F} - n) u^\mu \nabla_\mu F = 0,
\nonumber
\end{gather}
where we also used $u^\mu \nabla_\mu s = 0$ that comes from (\ref{s_equation}) and (\ref{equilibrium_fluid})
(recall that $n>0$ and $T>0$). If $u^\mu \nabla_\mu F = 0$, then, combined with our assumption 
(\ref{equilibrium_fluid}), we obtain (\ref{div_u_0}). 

Consider next the case where
$F \frac{\partial n}{\partial F} - n = 0$. This means that
\begin{gather}
n = a(s) F,
\nonumber
\end{gather}
for some function $a$ that depends only on $s$. 
In light of (\ref{first_law_2}),
\begin{gather}
\frac{\partial p}{\partial F} = n = a(s) F,
\nonumber
\end{gather}
which  implies upon integration that
\begin{gather}
p = \frac{1}{2} a(s) F^2 + b(s),
\label{p_a_b}
\end{gather}
where $b$ is a function of $s$ only. (\ref{def_enthalpy}) and (\ref{p_a_b}) combined now give
\begin{gather}
p - \varrho = -2 b(s).
\nonumber
\end{gather}
In particular this implies 
\begin{gather}
\left(\frac{\partial p}{\partial \varrho}\right)_s = 1,
\nonumber
\end{gather}
so that the fluid is stiff.

It remains the case where $F=0$. In this situation $p = -\varrho$ and we have a ``cosmological
constant fluid." Notice that both this last case and that of a stiff fluid are determined {\it a priori}
by the equation of state.

We conclude that the states of thermodynamic equilibrium of the Lichnerowicz's formulation
consist of the three possibilities (i) those characterized by (\ref{div_u_0}), which 
agree with the equilibrium states of Eckart's and MIS theory; and those where the
equations of motion reduce to those of a perfect fluid with an equation of state 
of either (ii) a stiff fluid or (iii) a cosmological constant fluid.

The possibilities (i) and (iii) include, in particular, the cases when $F$ is constant. 
More interestingly, we now show that: 

\begin{claim}
For an equation of state $p = w \varrho$, $w$ constant,  
if the specific enthalpy $F$ is constant, then generically (in a sense discussed below)
the fluid is  in 
thermodynamic equilibrium.
\end{claim}

This claim highlights the fundamental role played by the specific enthalpy in relativistic  fluids,
a point already made by Callen and Horowitz in the context of perfect fluids
\cite{CallenHorwitz}. The claim can in fact be shown for more general equations of state,
but we restrict ourselves to the case relevant to part \ref{part_applications} for the sake of brevity.

Assume, therefore, that  $p = w \varrho$ and $F$ is constant. Thus
\begin{gather}
dF = 0.
\label{dF_0}
\end{gather}
Write (\ref{def_enthalpy}) as
\begin{gather}
F = (1+z)\frac{p}{n},
\label{enthalpy_z}
\end{gather}
where $z = 1/w$. Then (\ref{dF_0}) and (\ref{enthalpy_z}) imply
\begin{gather}
d\left(\frac{p}{n}\right) = 0,
\nonumber
\end{gather}
which combined with (\ref{specific_enthalpy_2}) gives
\begin{gather}
d\epsilon = 0.
\nonumber
\end{gather}
Using $d\epsilon = 0$ in (\ref{first_law}) gives
\begin{gather}
T \, ds + \frac{p}{n^2} \,dn = 0,
\nonumber 
\end{gather}
or, after contracting with $u$,
\begin{gather}
T u^\mu \nabla_\mu s = -\frac{p}{n^2} u^\mu \nabla_\mu n.
\nonumber
\end{gather}
Using (\ref{rest_mass_eq}) to eliminate $u^\mu \nabla_\mu n$ yields
\begin{gather}
T u^\mu \nabla_\mu s = \frac{F}{1+z} \nabla_\mu u^\mu,
\label{relation_s_div_u}
\end{gather}
where we have also used (\ref{enthalpy_z}). Finally, using (\ref{relation_s_div_u})
to eliminate $s$ from (\ref{s_equation}) 
yields
\begin{gather}
F \nabla_\mu u^\mu \left[-\frac{n}{1+z} + \zeta \nabla_\mu u^\mu \right] = 0.
\label{nabla_mu_zero_eq}
\end{gather}
This equation has to hold for any choice of $z$ (i.e., $w$) and $\zeta$, 
 since it was derived without any restrictions on $w$ or $\zeta$,
thus typically 
in order for (\ref{nabla_mu_zero_eq}) to be satisfied one has to have
$\nabla_\mu u^\mu = 0$. This is the sense in which the claim holds generically: one has
to fine-tune the quantities in parenthesis in order to make it vanish. In fact, from the
point of view of the Cauchy problem discussed in section \ref{section_ENS_causal},
$n$, $\zeta$, and the first derivatives of $u$ are given Cauchy data, and thus we can choose them
so that $-\frac{n}{1+z} + \zeta \nabla_\mu u^\mu \neq 0$ at $t=0$, and thus also for small $t>0$ by
continuity. (\ref{nabla_mu_zero_eq}) then tells us that $\nabla_\mu u^\mu = 0$. This and the constancy 
of $F$, in turn, give (\ref{equilibrium_fluid}), as claimed.

From the above claim and the previous calculations it also follows that $u^\al \nabla_\al n = u^\al \nabla_\al p
= u^\al \nabla_\al \varrho = 0$, thus all these thermodynamic quantities are constant along 
the flow. 
\begin{remark}
Setting $\zeta = 0$ in (\ref{nabla_mu_zero_eq}), we see that in the case of a perfect fluid
the constancy of the enthalpy still gives $\nabla_\mu u^\mu = 0$, which in turn 
again implies $u^\al \nabla_\al n = u^\al \nabla_\al p
= u^\al \nabla_\al \varrho = 0$.
\end{remark}

\subsection{Entropy production}

We must verify that (\ref{second_law}) is compatible with the equations of motion derived
from (\ref{Lichnerowicz_stress_bulk}), i.e., with (\ref{div_T_eq}).
We shall restrict ourselves to the case (\ref{equation_state}).

Let us begin by deriving some useful identities.

Dividing (\ref{rho_equation_w}) by  $\varrho/F$ times (\ref{F_equation_w}) yields
\begin{gather}
\frac{ (1 - \zeta \frac{F}{\varrho} \nabla_\mu u^\mu) u^\al \nabla_\al \varrho }{
\frac{\varrho}{F}(1 - \zeta \frac{F}{\varrho} \nabla_\mu u^\mu) u^\al \nabla_\al  F }
= \frac{ (1+w) \varrho - 2\zeta F \nabla_\mu u^\mu}{ w \varrho - \zeta F \nabla_\mu u^\mu },
\nonumber
\end{gather}
provided that $\nabla_\mu u^\mu \neq  0$ and $u^\mu \nabla_\mu F \neq 0$. Notice that if one
of these two conditions does not hold then (\ref{second_law}) automatically follows. We can rewrite
the above as
\begin{gather}
 (1 - \zeta \frac{F}{\varrho} \nabla_\mu u^\mu) u^\al \nabla_\al \varrho 
= (1 + q) \frac{\varrho}{F}(1 - \zeta \frac{F}{\varrho} \nabla_\mu u^\mu) u^\al \nabla_\al  F,
\label{entropy_1}
\end{gather}
where
\begin{gather}
q = \frac{1 - \zeta \frac{F}{\varrho} \nabla_\mu u^\mu }{ w - \zeta \frac{F}{\varrho} \nabla_\mu u^\mu }.
\label{q_def}
\end{gather}
Contracting (\ref{first_law_2}) with $u$ and using (\ref{entropy_1}) gives
\begin{gather}
T u^\mu \nabla_\mu s = \frac{1 - q w}{1+w} u^\mu \nabla_\mu F,
\nonumber
\end{gather}
and then, after using (\ref{q_def}),
\begin{gather}
T u^\mu \nabla_\mu s = \frac{w - 1}{w+1} \frac{ \zeta \frac{F}{\varrho} \nabla_\mu u^\mu
u^\al \nabla_\al F }{
w -  \zeta \frac{F}{\varrho} \nabla_\mu u^\mu }.
\label{entropy_2}
\end{gather}
Using (\ref{def_enthalpy}) we can also write (\ref{entropy_2}) as 
\begin{gather}
T u^\mu \nabla_\mu s = \frac{w - 1}{w+1} \frac{ \zeta  \nabla_\mu u^\mu
u^\al \nabla_\al F }{
 \frac{w}{1+w} n -  \zeta \nabla_\mu u^\mu }.
\label{entropy_3}
\end{gather}

We shall show that:
\begin{claim}
Under assumption (\ref{equation_state}), the following are 
sufficient conditions for (\ref{second_law}) to hold:
\begin{align}
& (i)\, 0 < w \leq 1 \, \text{ and } \, 1 - \zeta \frac{F}{\varrho} \nabla_\mu u^\mu > 0;
\nonumber\\
& (ii)\,  -1 < w \leq 1 \, \text{ and } \, \frac{w}{1+w} n - \zeta \nabla_\mu u^\mu > 0;
\nonumber \\
& (iii)\, w < -1 \, \text{ and } \frac{w}{1+w} n - \zeta \nabla_\mu u^\mu < 0.
\nonumber
\end{align}
\label{claim_entropy_production}
\end{claim}
At this moment the above conditions may seem arbitrary, but they are connected
with relevant scenarios in the study of causality and well-posedness of the Einstein-Navier-Stokes
system discussed in section \ref{section_ENS_causal} and applications to cosmology in 
part \ref{part_applications}. In view of (\ref{rest_mass_eq}), conditions (ii) and (iii) can be translated
into conditions on the evolution of $n$. Several other sufficient conditions can be derived, but we believe
that without a specific choice of (\ref{zeta_function}) or a particular application in mind, many of these conditions
will look artificial. Notice also that if $w=-1$, then $F=0$ and the second law is automatically satisfied.

\begin{remark}
Condition (i) above says that we can guarantee that the second law is satisfied if we take $\zeta$ sufficiently small.
This is in accordance with the intuitive idea that for $\zeta$ small, our system should be a perturbation of a perfect fluid
 and, therefore, we would expect the second law to hold (since it does for perfect fluids). This is not, however, patent
 directly from (\ref{rho_equation}) and (\ref{first_law_2}), since $\zeta$ appears as an overall factor on
 the viscous terms in (\ref{rho_equation}). The point is that $u^\mu \nabla_\mu F$ and $\nabla_\mu u^\mu$ (which
 of course depend  on $\zeta$) need not to scale  with $\zeta$ in the same way, but in order to show this we need to write the
 equations in a particular way.
\end{remark}

Because of (\ref{s_equation}), if $\nabla_\mu u^\mu$ and $u^\mu \nabla_\mu F$ have equal signs then
(\ref{second_law}) holds. Thus, it suffices to investigate the case where 
$\nabla_\mu u^\mu$ and $u^\mu \nabla_\mu F$ have opposite signs.

Let us start with (i), so assume the corresponding statements. 
We need to verify that $ w -  \zeta \frac{F}{\varrho} \nabla_\mu u^\mu > 0$. Assume first that
$\nabla_\mu u^\mu >0$, so that $u^\mu \nabla_\mu F < 0$. Then (\ref{F_equation_w}) gives
\begin{gather}
w \nabla_\mu u^\mu - \zeta \frac{F}{\varrho} (\nabla_\mu u^\mu)^2 > 0,
\nonumber
\end{gather}
or 
\begin{gather}
 ( w - \zeta \frac{F}{\varrho} \nabla_\mu u^\mu ) \nabla_\mu u^\mu > 0.
\nonumber
\end{gather}
This implies $w - \zeta \frac{F}{\varrho} \nabla_\mu u^\mu > 0$ since $\nabla_\mu u^\mu$ is
 positive. Similarly, if we assume $\nabla_\mu u^\mu < 0$ and $u^\mu \nabla_\mu F >0$ and invoke (\ref{F_equation_w}) 
we conclude
\begin{gather}
 ( w - \zeta \frac{F}{\varrho} \nabla_\mu u^\mu ) \nabla_\mu u^\mu < 0,
\nonumber
\end{gather}
implying $w - \zeta{F}{\varrho} \nabla_\mu u^\mu > 0$ as now $\nabla_\mu u^\mu$ is negative. This shows (i).

(ii) and (iii) follow directly from (\ref{entropy_3}).

\subsection{Energy conditions}

The constraints imposed on the matter fields for 
(\ref{Lichnerowicz_stress_bulk})
 by energy conditions are readily obtained from
the known constraints for (\ref{perfect_stress}) after redefining
$p \mapsto p - \zeta \nabla_\mu C^\mu$. We shall, however, give an argument from scratch for the 
reader's convenience. Recall that a stress-energy tensor $T_{\al\be}$
is said to satisfy the \textbf{weak energy condition} if 
$T_{\al\be} \xi^\al\xi^\be \geq 0 $ for all time-like vectors $\xi^\al$; the \textbf{null-energy condition}
if $T_{\al\be} N^\al N^\be \geq 0$ for all future-pointing null vectors $N^\al$; the \textbf{strong energy
condition} if $T_{\al\be} \xi^\al \xi^\be - \frac{1}{2} T \xi^\al \xi_\al \geq 0$ 
for all time-like vectors $\xi^\al$, where $T$ is the trace of $T_{\al\be}$; the 
\textbf{dominant energy condition} if $-T_{\al\be} \xi^\al $ is a future time-like or null vector for 
all future directed time-like vectors $\xi^\al$. We notice that in the  time-like case
 it suffices to consider normalized time-like vectors.

Consider an orthonormal frame $\{ e_A \}_{A=0}^3$ such that $e_0 = u$, where we  use upper
Latin letters to number the frames. Let $\eta_{AB} = \operatorname{diag}(-1,1,1,1) = \eta^{AB}$, so that
\begin{gather}
e^\al_A e^\be_B g_{\al\be} = \eta_{AB},
\nonumber
\end{gather}
\begin{gather}
g^{\al\be} =  e^\al_A e^\be_B \eta^{AB},
\nonumber
\end{gather}
where $e_A^\al$ is the $\al$-coordinate of $e_A$ in the $x^\al$-coordinates (see, e.g.,  
\cite{FriRenCauchy,WaldBookGR1984}
 for a more detailed presentation of the frame formalism, or 
 \cite{DisconziRemarksEinsteinEuler} for a brief discussion).
We can further assume, without loss of generality, that at the origin of the coordinate system
the metric $g_{\al\be}$ is also given by $\operatorname{diag}(-1,1,1,1)$. An arbitrary
future directed (normalized) time-like vector
can be written
\begin{gather}
\xi^\al = \Gamma (e_0^\al + a e_1^\al + b e_2^\al + c e_3^\al ),
\nonumber
\end{gather}
with $a^2 + b^2 +c^2 < 1$ and 
\begin{gather}
\Gamma = \frac{1}{ \sqrt{ 1- a^2 - b^2 -c^2 }},
\nonumber
\end{gather}
whereas an arbitrary future directed null-like vector reads
\begin{gather}
N^\al = e_0^\al + a e_1^\al + b e_2^\al + c e_3^\al ,
\nonumber
\end{gather}
with $a^2 + b^2 +c^2 = 1$.
From the above relations and (\ref{Lichnerowicz_stress_bulk}) one readily computes
\begin{gather}
T_{\al\be} \xi^\al \xi^\be = \Gamma^2[ \varrho +  (a^2 + b^2 +c^2)p - \zeta
(a^2 + b^2 +c^2 ) \nabla_\mu C^\mu ) ],
\label{for_weak_energy}
\end{gather}
\begin{gather}
T_{\al\be} N^\al N^\be = p + \varrho - \bv \nabla_\mu C^\mu, 
\label{for_null_energy}
\end{gather}
\begin{gather}
T_\al^\al = (3p - \varrho) -3 \zeta \nabla_\mu C^\mu,
\label{for_strong_energy}
\end{gather}
and
\begin{gather}
-T_{\al\be} \xi^\be = 
\Gamma \varrho e_{0\al} - \Gamma p   ( a e_{1\al} + b e_{2\al} + c e_{3\al} )
+ \Gamma  ( a e_{1\al} + b e_{2\al} + c e_{3\al} )\nabla_\mu C^\mu,
\label{for_dominant_energy}
\end{gather}
where $e_{A\al}=g_{\al\be} e^\be_A $. 

Imposing each one of the energy conditions stated above, we obtain different restrictions for the matter
fields, as follows. Setting first $a=b=c=0$ in (\ref{for_weak_energy}) we get $\varrho \geq 0$, then using
$a^2 + b^2+c^2 < 1 $  gives
\begin{gather}
\zeta( a^2 + b^2+c^2 ) \nabla_\mu C^\mu \leq \varrho + 
(a^2 + b^2+c^2) p < p + \varrho,
\nonumber
\end{gather}
for $p\geq 0$.  Hence $z \zeta \nabla_\mu C^\mu < p + \varrho$ for all $0 \leq z < 1$.
Similarlly (\ref{for_null_energy}) gives
$\zeta \nabla_\mu C^\mu \leq  p + \varrho$.

Combining (\ref{for_weak_energy}) and (\ref{for_strong_energy}) with $a=b=c=0$ gives
$3\zeta \nabla_\mu C^\mu \leq \varrho +3p$. Using this 
with (\ref{for_weak_energy}) and (\ref{for_strong_energy})  but now with $b=c=0$ produces
$(3-a^2) \zeta \nabla_\mu C^\mu \leq a^2(\varrho - p) + 3p + \varrho$, which can be written
$(3-z) \zeta \nabla_\mu C^\mu \leq (3-z) p + (1+z) \varrho$, for all $0\leq z \leq 1$.

The dominant energy condition gives $g^{\al\be}(-T_{\al\la}\xi^\la)(-T_{\be \tau} \xi^\tau) \leq 0$.
Combining with (\ref{for_dominant_energy}) and setting $a=b=c=0$ we get $\varrho^2 \geq 0$. 
And if we set only $b=c=0$, then we obtain $\varrho^2 \geq a^2 (p-\zeta \nabla_\mu C^\mu)^2$,
which can be written as $ |p - \zeta \nabla_\mu C^\mu | \leq \varrho$.

We now summarize the results:

\begin{claim}
For the stress-energy tensor (\ref{Lichnerowicz_stress_bulk}), the energy conditions read as follows.

\vskip 0.1cm
\noindent Weak energy condition:
\begin{gather}
\varrho \geq 0 \, \text{ and } \, z \zeta \nabla_\mu C^\mu < p + \varrho \,
\text{ for all } \, 0 \leq z < 1.
\nonumber
\end{gather}
\noindent Null energy condition:
\begin{gather}
 \zeta \nabla_\mu C^\mu \leq p + \varrho. 
\nonumber
\end{gather}
\noindent Strong energy condition:
\begin{gather}
3\zeta \nabla_\mu C^\mu \leq \varrho + 3p  \, \text{ and } \, 
(3-z) \zeta \nabla_\mu C^\mu \leq (3-z) p + (1+z) \varrho \,
\text{ for all } \, 0 \leq z < 1.
\nonumber
\end{gather}
\noindent Dominant energy condition:
\begin{gather}
\varrho \geq 0 \, \text{ and } \, |p - \zeta \nabla_\mu C^\mu | \leq \varrho.
\nonumber
\end{gather}

\end{claim}

\subsection{The non-relativistic limit\label{section_nr_limit}}

By the non-relativistic limit we mean regimes where the speed $|v|$ of the fluid, as measured by an Eulerian
observer (see below), is small compared to $c$,  
where $|v|$ is the norm of the 3-velocity $v$ in the metric $\overline{g}$
induced on $\{t=\text{constant}\}$ hypersurfaces:
\begin{gather} 
\frac{|v|}{c} \ll 1.
\nonumber
\end{gather}
It is also assumed that 
the energy density of the fluid is given essentially by its rest-mass density, 
and that the pressure contribution to the energy density is negligible, so
\begin{gather}
\epsilon \ll c^2,
\nonumber
\end{gather}
and
\begin{gather}
\frac{p}{n} \ll c^2.
\nonumber
\end{gather}
Finally, in the non-relativistic limit, the three-metric $\overline{g}_{ij}$ 
converges to the Euclidean metric:
\begin{gather}
\overline{g}_{ij} \rar \de_{ij}.
\nonumber
\end{gather}
In order to carry out the non-relativistic limit, it is convenient to reinstate $c$ in all the equations.
Then  (\ref{density_mass_internal_energy}) reads
\begin{gather}
\varrho =c^2 n ( 1 + \frac{\epsilon}{c^2} ),
\nonumber
\end{gather}
so that 
\begin{gather}
p + \varrho = c^2n( 1 + \frac{\epsilon}{c^2} + \frac{p}{c^2 n} ).
\label{p_plus_rho_c}
\end{gather}
We see from (\ref{p_plus_rho_c}) that $F \rar c^2$ in the non-relativistic limit, or $F\rar 1$ in 
units where $c=1$. From this it follows that the non-relativistic limits of Lichnerowicz
and Eckart's stress-energy tensors are the same. Thus it suffices  to calculate the 
non-relativistic limit obtained from (\ref{Eckart_stress}), which we shall do since
the equations derived from (\ref{Eckart_stress}) are simpler than those obtained from 
Lichnerowicz's formulation. The non-relativistic limit of Eckart's theory is carried
out, for instance, in \cite{RezzollaZanottiBookRelHydro}, but we shall present the argument 
for the reader's convenience. In this section we shall also consider a non-zero shear viscosity
so that the full Navier-Stokes equations are obtained in the limit (although we still take $\hc = 0$).

Choose coordinates where the metric is written in as
\begin{gather}
g = - (dx^0)^2 + g_{ij} dx^i dx^j,
\label{metric_decomposion}
\end{gather}
so that for any vector $X$
\begin{gather}
X_0 = g_{0\al} X^0 = - X^0,
\nonumber
\end{gather}
and spatial indices can be raised and lowered with the spatial Riemannian metric 
$\overline{g} = g_{ij} dx^i dx^j$,
\begin{gather}
X_i = g_{i\al} X^\al = g_{ij} X^j.
\nonumber
\end{gather}
The 4-velocity measured by an Eulerian observer is defined by
\begin{gather}
u^\al =  \ga (1, \frac{v^i}{c}), 
\label{definition_Eulerian_v}
\end{gather}
where
\begin{gather}
\ga = \frac{1}{ \sqrt{ 1 - \frac{|v|^2}{c^2} } }.
\nonumber
\end{gather}
Physically\footnote{For the definition of $v$ when the metric does not take the form (\ref{metric_decomposion}), see section 7.1 of \cite{RezzollaZanottiBookRelHydro}.
See section III.6 of  \cite{ZeeEinsteinNutshell} for more discussion on the physical meaning of $v$.}, $v$ is the velocity actually measured by a stationary observer, so 
one always has $|v| \leq c$. From (\ref{definition_Eulerian_v}),
\begin{gather}
u_\al =  \ga (-1, \frac{v_i}{c}), 
\nonumber
\end{gather}
so that 
\begin{gather}
u^\al = (1 + O\left(\frac{|v|^2}{c^2}\right), \frac{v^i}{c} + O\left(\frac{|v|^3}{c^3}\right) \, ),
\label{expansion_u_up}
\end{gather}
and
\begin{gather}
u_\al = (-1 + O\left(\frac{|v|^2}{c^2}\right), \frac{v_i}{c} + O\left(\frac{|v|^3}{c^3}\right) \, ),
\label{expansion_u_down}
\end{gather}
We shall also make the time dependence explicit upon recalling that
\begin{gather}
x^0 = ct.
\label{x_zero_t}
\end{gather}
The momentum equation $\pi^{\al\be}\nabla^\mu T^E_{\mu\be} = 0$ can be written as
\begin{align}
\begin{split}
(p + \varrho) u^\mu \nabla_\mu u_\al + \pi_\al^\mu \nabla_\mu p - \nabla_\al (\bv
\nabla_\mu u^\mu ) 
- u^\mu \nabla_\mu ( \bv \nabla_\la u^\la u_\al)
- 2 \nabla_\mu( \sv \si^\mu_\al ) +
2 \sv u_\al \si^{\mu\nu}\si_{\mu\nu} = 0,
\end{split}
\nonumber
\end{align}
where
\begin{gather}
2 \si_{\al\be} = \nabla_\al u_\be + \nabla_\be u_\al + u_\al u^\mu \nabla_\mu u_\be + u_\be u^\mu \nabla_\mu u_\al
-\frac{2}{3} \pi_{\al\be} \nabla_\mu u^\mu.
\label{two_shear_def}
\end{gather}
Or, upon restoring $c$, (\ref{Eckart_stress}) reads,
\begin{gather}
T^E_{\al \be} = (p+\varrho) u_\al u_\be + pg_{\al\be} 
- c ( \bv - \frac{2}{3}  \sv ) \pi_{\al\be} \nabla_\mu u^\mu - 
c\sv 
\pi_\al^\mu \pi_\be^\nu(\nabla_\mu u_\nu + \nabla_\nu u_\mu).
\nonumber
\end{gather}
so that
\begin{align}
\begin{split}
(p + \varrho) u^\mu \nabla_\mu u_\al + &  \pi_\al^\mu \nabla_\mu p - c\nabla_\al (\bv
\nabla_\mu u^\mu ) 
- c u^\mu \nabla_\mu ( \bv \nabla_\la u^\la u_\al) \\
&
- c \nabla_\mu( \sv \si^\mu_\al ) +
c \sv u_\al \si^{\mu\nu}\si_{\mu\nu} = 0.
\end{split}
\label{relativistic_NS_equation_c}
\end{align}
Our goal is to compute each term in (\ref{relativistic_NS_equation_c}) using
(\ref{p_plus_rho_c}),
 (\ref{metric_decomposion}),
(\ref{expansion_u_up}),  (\ref{expansion_u_down}), and (\ref{x_zero_t}).  We begin with
\begin{gather}
\pi_{00} = O\left(\frac{|v|^2}{c^2}\right),
\nonumber
\end{gather}
\begin{gather}
\pi_{0i} = -\frac{v_i}{c} + O\left(\frac{|v|^3}{c^3}\right),
\nonumber
\end{gather}
and 
\begin{gather}
\pi_{ij} = g_{ij} + O\left(\frac{|v|^2}{c^2}\right).
\nonumber
\end{gather}
We have
\begin{gather}
u^\mu   \nabla_\mu  u_\al = \frac{1}{c}(\nabla_t u_\al + v^i \nabla_i u_\al ) 
+  O\left(\frac{|v|^3}{c^3}\right).
\nonumber
\end{gather}
Setting $\al = 0$ and $\al = j$ we find, respectively
\begin{gather}
u^\mu \nabla_\mu u_0 =  O\left(\frac{|v|^2}{c^3}\right),
\nonumber
\end{gather}
and
\begin{gather}
u^\mu \nabla_\mu u_j = \frac{1}{c^2}(\nabla_t v_j + v^i \nabla_i v_j ) 
+  O\left(\frac{|v|^3}{c^3}\right).
\nonumber
\end{gather}
Similarly,
\begin{gather}
\nabla_\mu u^\mu = \frac{1}{c} \nabla_i v^i + O\left(\frac{|v|^2}{c^3}\right).
\nonumber
\end{gather}
so that
\begin{gather}
c \nabla_0 (\bv \nabla_\mu u^\mu ) =  \frac{1}{c} \nabla_t( \bv \nabla_i v^i )+ O\left(\frac{|v|^2}{c^2}\right)
= O\left(\frac{|v|}{c}\right),
\nonumber
\end{gather}
and
\begin{gather}
c \nabla_j (\bv \nabla_\mu u^\mu ) =  \nabla_j( \bv \nabla_i v^i ) +
O\left(\frac{|v|^2}{c^2}\right),
\nonumber
\end{gather}
and also
\begin{gather}
c u^\mu \nabla_\mu ( \bv \nabla_\la u^\la u_\al) =  O\left(\frac{|v|}{c}\right).
\nonumber
\end{gather}
From (\ref{two_shear_def}), we obtain
\begin{gather}
2 \si_{00} = O\left(\frac{|v|^2}{c^3}\right),
\nonumber
\end{gather}
\begin{gather}
2 \si_{0i} = O\left(\frac{|v|}{c^2}\right),
\nonumber
\end{gather}
and
\begin{gather}
2 \si_{ij} = \frac{1}{c} (\nabla_i v_j + \nabla_j v_i - \frac{2}{3} g_{ij} \nabla_k v^k ) + O\left(\frac{|v|^2}{c^2}\right).
\nonumber
\end{gather}
We can thus compute
\begin{gather}
2c \nabla_\mu (\sv \si_0^\mu ) = O\left(\frac{|v|}{c}\right),
\nonumber
\end{gather}
and
\begin{gather}
2c \nabla_\mu (\sv \si^\mu_j ) = \nabla_i
( \sv ( \nabla^i v_j + \nabla_j v^i - \frac{2}{3} g^i_j \nabla_k v^k ) ) 
+ O\left(\frac{|v|}{c}\right).
\nonumber
\end{gather}
Combining all of the above, we finally obtain
\begin{align}
\begin{split}
&
  n ( \nabla_t v_j + v^i \nabla_i v_j ) + 
\nabla_j p 
-  \nabla^i [ 
\sv ( \nabla_i v_j + \nabla_j v_i - \frac{2}{3} g_{ij} \nabla_k v^k ) 
+ \bv \nabla_k v^k g_{ij} ] \\
& + 
O\left(\frac{|v|}{c}\right) + O\left(\frac{\epsilon}{c^2}\right)
+ O\left(\frac{p}{n c^2}\right)
= 0.
\end{split}
\nonumber
\end{align}
Taking
\begin{gather} 
\frac{|v|}{c} \rar 0, \, 
 \frac{\epsilon}{c^2} \rar 0, \frac{p}{n} \rar 0,
 \, \text{ and } \, g_{ij} \rar \de_{ij},
\nonumber
\end{gather}
we arrive at (the momentum equation of) the Navier-Stokes equations:
\begin{align}
\begin{split}
  \partial_t v_j + v^i \partial_i v_j  + 
\frac{1}{n} \partial_j p 
- \frac{1}{n} \partial^i [ 
\sv ( \partial_i v_j + \partial_j v_i - \frac{2}{3} \de_{ij} \partial_k v^k ) 
+ \bv \partial_k v^k\de_{ij} ]
= 0.
\end{split}
\label{NS_limit_general}
\end{align}
Notice that in this limit $n$ is re-interpreted as the mass-density in the non-relativistic Navier-Stokes equations
(\ref{NS_limit_general}). This is consistent with $n$ being, in relativity,  the rest-mass density measured 
by a co-moving observer.

Similarly it can be easily shown that the non-relativistic limit of $u^\be \nabla^\al T^E_{\al\be} = 0$ 
and (\ref{rest_mass_eq}) give, respectively, the heat-conduction 
and continuity equations of non-relativistic physics. We refer the reader to
\cite{RezzollaZanottiBookRelHydro} for details.

If we now turn to the more familiar situation of an incompressible fluid, we can  set $n=1$ and 
$\partial_k v^k = 0$ in (\ref{NS_limit_general}), obtaining
\begin{align}
\begin{split}
\frac{\partial v_j}{\partial t} + v^i \partial_i v_j + 
 \partial_j p - \sv \partial^i  \partial_i v_j
 = 0,
\end{split}
\label{NS_limit_incompressible}
\end{align}
where we assumed that $\sv$ is constant for simplicity. (\ref{NS_limit_incompressible})
is of course the familiar form of the incompressible Navier-Stokes equations.

\subsection{A second order formulation?}

In this section, we point out how the ideas of second order theories briefly discussed
in section \ref{section_theories} could be incorporated in the model here studied. More precisely,
we consider a second order formulation such that the Einstein-Navier-Stokes system 
is obtained as a particular limit. For simplicity we shall discuss only the case where bulk viscosity
is present. Thus, we consider an entropy current of the form (\ref{entropy_current_second_order})
with coefficients $\be_1 = \be_2 = \al_0 = \al_1 = 0$,  and suppose that
\begin{gather}
\Pi = -\bv  f \nabla_\mu C^\mu,
\label{new_pi}
\end{gather}
where $f$ is a variable to be determined. (\ref{Lichnerowicz_stress_bulk}) and its corresponding first
order formulation are then obtained as the particular case $\be_0 = 0$ and $f = 1$. With these
definitions, one finds
\begin{align}
\begin{split}
T \nabla_\mu S^\mu = \bv f \nabla_\mu C^\mu( \nabla_\mu u^\mu +
\be_0[ 
u^\mu \nabla_\mu \bv f \nabla_\la C^\la + 
\bv u^\mu \nabla_\mu f \nabla_\la C^\la + 
\bv f u^\mu \nabla_\mu \nabla_\la C^\la ) ].
\end{split}
\nonumber
\end{align}
According to the philosophy of second order theories outlined in section \ref{section_theories},
we now postulate an evolution equation for $f$ that ensures that $\nabla_\mu S^\mu  \geq 0$.
The simplest such choice is
\begin{align}
\begin{split}
\bv \nabla_\la C^\la u^\mu \nabla_\mu f 
+ \bv f u^\mu \nabla_\mu \nabla_\la C^\la
+
fu^\mu \nabla_\mu \bv  \nabla_\la C^\la 
= \frac{1}{\be_0} \bv f \nabla_\mu C^\mu - \frac{1}{\be_0} \nabla_\mu u^\mu.
\end{split}
\label{f_eq_second_order}
\end{align}
Equation (\ref{f_eq_second_order}) is now appended to the previous equations of motion,
and in this new formulation, the system describing a relativistic viscous fluid consists
of equations  (\ref{rest_mass_eq}), (\ref{Einstein_eq}), (\ref{rho_equation}), and (\ref{u_equation}) suitably
modified according to (\ref{new_pi}) (which replaces the term in $-\bv \nabla_\mu C^\mu$ in 
(\ref{Lichnerowicz_stress_bulk})), and equation 
(\ref{f_eq_second_order}). These equations are, as before, 
supplemented by the relations 
,  (\ref{normalization_u}),  (\ref{first_law}),  (\ref{equation_state_general}) and (\ref{zeta_function}). 
In this situation, the second law of thermodynamics is automatically satisfied. On the other
hand, the equations of motion, which were already complicated before the introduction
of $f$, become highly complex. It is not clear at this point what can be said about this new
system of equations, in particular if it is causal. On the other hand, these new equations
are based on a formulation that has been shown to be causal, under certain assumptions, as
a first order theory. If we subscribe to the (not totally unfounded) view that when one promotes a first order 
theory to a second order one, the equations behave ``better," one can hope that this new
Einstein-Navier-Stokes system will lead to causal equations of motion.

\section{Lichnerowicz's stress energy tensor and its motivation\label{section_motivation}}

In this section we discuss motivations for considering \ref{Lichnerowicz_stress}. First, we 
recall some of the assumptions made in the definition of the four-velocity $u$, following mainly the
discussion in \cite{RezzollaZanottiBookRelHydro}.
Then, we briefly review
the prominent role played by the enthalpy current (\ref{def_C}) in relativistic fluids. Then, we give the
original argument of Lichnerowicz \cite{LichnerowiczBookGR}. Finally, we advance further arguments
that motivate (\ref{Lichnerowicz_stress}).

\subsection{The velocity of a relativistic viscous fluid\label{section_velocity_fluid}}

As pointed out in the introduction, the extension of the classical (non-relativistic) stress-energy
tensor of the Navier-Stokes equations to general relativity is ambiguous. But even a choice of a stress-energy
tensor does not entirely rule out the arbitrariness in the choices of  fluid variables. To see this, consider
first the case of a perfect fluid. One can then define a frame which is instantaneously moving with the fluid by
considering the velocity of fluid particles. The definition of such a frame, however, is ambiguous
when viscosity and heat conduction are present. This is because the rest-mass density current, the entropy 
current, and the stress-energy momentum tensor should be considered as the primary physical variables,
with their relations to the four-velocity determined from further considerations that take into account
the equations of motion and the second law of thermodynamics.

If we denote the rest-mass density current by $n^\mu$, we see that there are two natural ways to define the 
four-velocity. One would be to take it to be related to $n^\mu$ via $n^\mu = n u^\mu$; the other would be
to consider an eigenvector of the stress-energy tensor, i.e., $u^\mu = T^\mu_\nu u^\nu$. In both cases,
it is assumed $u^\mu u_\mu = -1$. If we denote the first choice by $u_E$ and the second by $u_L$,
it holds that \cite{MIS-2}
\begin{gather}
u_E = u_L + f(\bv, \sv, \hc),
\nonumber
\end{gather}
where $f$ is a function of the bulk and shear viscosities and the heat conduction that vanishes
when these quantities are zero. Thus in particular the two definitions of $u$ agree for a perfect fluid,
but in general one has to make a choice. The first choice, $u_E$, is known as the Eckart frame
and the second one, $u_L$, as the Landau frame. From   (\ref{Lichnerowicz_stress}) and (\ref{rest_mass_eq}) 
it is seen that we have adopted the Eckart frame in this work. While such a choice is widely used in the 
literature of relativistic viscous fluids, one has to be aware that it is a choice, but one which nonetheless
does not seem {\it a priori} better motivated than the use of $u_L$. We refer the reader to \cite{RezzollaZanottiBookRelHydro, Weinberg_GR_book} for more details.

\subsection{The enthalpy current\label{section_enthalpy_current}} 

The dynamic velocity or enthalpy current defined
by (\ref{def_C}) plays a fundamental role in the theory of both perfect and viscous fluids. 
Perhaps its most important feature is that the vorticity of a relativistic (perfect or viscous)
fluid is defined in terms of $C$ and not $u$ (recall definition (\ref{vorticity})).
The reason for this is as follows. The definition of the vorticity $\Om$ 
is intended to quantify the formation of eddies in a fluid, and the number of such eddies on
a surface enclosed by a closed loop should remain constant in time (this is one of the interpretations
of the Kelvin circulation theorem in classical physics). It turns out that for the equations of motion 
to imply this conservation property one has to define the vorticity as in (\ref{vorticity}), i.e., in terms
of $C$. The reader is referred to \cite{LichnerowiczBookGR,RezzollaZanottiBookRelHydro}
for proofs of these statements\footnote{One sometimes defines the so-called kinematic vorticity
as $\nabla_\mu u_\nu - \nabla_\nu u_\mu$, thus solely in terms of $u$. As the definition implies, this quantity, 
is purely knematic and does not capture the dynamics of vortices in the fluid.}.

Another interesting property of the enthalpy current is the following. In classical fluids, the incompressibility 
of a fluid is given by $\partial_i v^i = 0$, where $v$ is the fluid velocity (defined only in three dimensions in this
case). The relativistic analogue of an incompressible fluid is a stiff fluid, i.e., one  
where the speed of propagation of sound waves equals the speed of light. It is natural to ask if the stiffness
of a relativistic fluid can be characterized by the divergence of a quantity in the very same fashion
that it is characterized by $\partial_i v^i = 0$ in the case of classical fluids.
In the case of a perfect fluid, a calculation similar to that of section \ref{section_equilibrium} shows that 
this is indeed the case if we consider the dynamic velocity   rather than the four-velocity, i.e.,
if a perfect\footnote{The situation when viscosity is present is more complicated, but we can still
relate stiffness to the divergence of $C$. An example of this situation is given in section \ref{section_equilibrium}.} fluid is stiff, then
\begin{gather}
\nabla_\mu C^\mu = 0,
\nonumber
\end{gather}
leading therefore to a generalization of the the incompressibility notion of a classical fluid.

At least two further features of $C$ should be stressed. The current $C$ is, by definition, naturally
linked to the flow of the enthalpy $F$ in space-time. The enthalpy, in turn, encodes many
important dynamic features of the fluid. For instance, it is tied to the fluid's thermodynamic equilibrium, as
seen in section \ref{section_equilibrium}. For perfect fluids, the enthalpy is also responsible for
connecting the intrinsic geometry of the space-time to the fluid flow lines, as follows. If we consider
the conformal metric 
\begin{gather}
\widetilde{g} = F^2 g,
\label{conformal_metric}
\end{gather}
(assuming $F>0$) then it is possible to show that the momentum equation
$\pi^{\al\be}\nabla_\ga T^\ga_\be = 0$ corresponds to geodesics of the metric $\widetilde{g}$
\cite{Lichnerowicz_MHD_book}. Further properties of perfect fluids that are essentially 
controlled by the enthalpy have been explored by Callen and Horwitz \cite{CallenHorwitz},
and also used by Christodoulou \cite{ChristodoulouShocks}.

In summary, these observations suggest that, at a conceptual level, one should consider
the enthalpy current rather than the four-velocity as a more fundamental quantity. This point
of view is not in conflict with any of the known properties of fluids as we can switch back and forth
between $u$ and $C$ via (\ref{def_C}). Nor does it imply any difficulties with established properties
of non-relativistic fluids since $F \rar 1$ in the non-relativistic limit (see section \ref{section_nr_limit}).

\subsection{Lichnerowicz's construction}
In this section, we present the argument given by Lichnerowicz that motivated (\ref{Lichnerowicz_stress}).
This argument was given in \cite{LichnerowiczBookGR}, which we follow closely.

The starting point relys on two principles. First, as remarked above, the metric $\widetilde{g}$ given by
(\ref{conformal_metric}) leads to a fluid flow generated by geodesics when dissipation is absent. Hence,
one imagines that $\widetilde{g}$ is the correct metric that generalizes to relativity the purely 
kinematic properties that one encounters in classical fluids. Second, $C$ should be thought of
as a primary variable in light of the observations of section \ref{section_enthalpy_current}.
Once these two principles are adopted, it is natural to consider the quantity $\widetilde{C}$
given by
\begin{gather}
\widetilde{C}_\al = C_\al 
\nonumber
\end{gather}
and
\begin{gather}
\widetilde{C}^\al = \widetilde{g}^{\al\be} C_\be 
\nonumber
\end{gather}
since $\widetilde{C}$ is now unitary in the metric $\widetilde{g}$,
\begin{gather}
\widetilde{g}^{\al\be} \widetilde{C}_\al \widetilde{C}_\be = \widetilde{C}^\al \widetilde{C}_\al = - 1,
\label{normalization_C_tilde}
\end{gather}
provided that $C^\al C_\al = -F^2$. Hence, $\widetilde{C}$ is normalized (in the metric $\widetilde{g}$)
similarly to the normalization (\ref{normalization_u})  of $u$ (in the metric $g$). Notice that 
$\widetilde{C}^\al$ can also be written as $\widetilde{C}^\al = F^{-1} u^\al$. 

With these constructions, one can consider a frame where
\begin{gather}
\widetilde{C}^0 = 1, \,  \widetilde{C}^i = 0,
\label{coord_C_tilde}
\end{gather}
(notice that such a frame is not the same in which $u^0 = 1$) and define, similarly to what is done in classical physics, a tensor of stresses of the form
\begin{gather}
P_{ij} = P \widetilde{g}_{ij} - \mu ( \widetilde{\nabla}_i \widetilde{C}_j +  \widetilde{\nabla}_j 
\widetilde{C}_i ),
\label{def_P_tensor}
\end{gather}
where $\mu$ is a viscosity coefficient, $P$ a scalar associated with the pressure (including both
perfect and viscous effects associated with the pressure), and $\widetilde{\nabla}$ the covariant
derivative of the metric $\widetilde{g}$. Notice that this
expression differs from the analogous one for classical fluids by terms of order 
$1/c^2$ (see (\ref{p_plus_rho_c})). The stress energy tensor
that we seek to define is therefore written
\begin{gather}
T_{\al\be} = \varrho u_\al u_\be + P_{\al\be},
\nonumber
\end{gather}
and if we adopt the standard practice that $\varrho$ is indeed the energy density measured
at a co-moving frame we have $u^\be P_{\al \be} = 0$, so that in coordinates
(\ref{coord_C_tilde})
 we can write
\begin{gather}
P_{i0} = 0.
\label{P_coord_C_tilde}
\end{gather}
To write (\ref{def_P_tensor}) in arbitrary coordinates, notice first that $\widetilde{g}_{ij}$ in (\ref{def_P_tensor}) is exactly the metric induced on the space orthogonal to $\widetilde{C}$, which in arbitrary coordinates is
written as
\begin{gather}
\widetilde{\pi}_{\al\be}=\widetilde{g}_{\al\be} + \widetilde{C}_\al \widetilde{C}_\be.
 \label{LichnerowiczProjector}
\end{gather}
We find 
\begin{gather}
\widetilde{\pi}_{\al\be}\widetilde{C}^\be = 0
 \label{LichnerowiczProjectorXC}
\end{gather}
in analogy with (\ref{projection_u}).
Next, because of  (\ref{normalization_C_tilde}), we have
\begin{gather}
\widetilde{C}^\al \widetilde{\nabla}_\be \widetilde{C}_\al = 0.
\label{derivative_C_tilde_zero}
\end{gather}
Thus, we can verify that in light of (\ref{P_coord_C_tilde}) and (\ref{derivative_C_tilde_zero}),
 the term $ ( \widetilde{\nabla}_i \widetilde{C}_j +  \widetilde{\nabla}_j 
\widetilde{C}_i )$ in (\ref{def_P_tensor}) assumes the following form in general coordinates:
\begin{gather}
 \widetilde{\nabla}_\al \widetilde{C}_\be +
  \widetilde{\nabla}_\be \widetilde{C}_\al +
\widetilde{C}^\la(  \widetilde{\nabla}_\la \widetilde{C}_\al \widetilde{C}_\be + 
  \widetilde{\nabla}_\la \widetilde{C}_\be \widetilde{C}_\al  ).
  \nonumber
\end{gather}
Next, we need to postulate a form for $P$. Since $T_{\al\be}$ has to reduce to the stress-energy tensor
of a perfect fluid in the absence of dissipation, we must have
\begin{gather}
P = F^{-2} p +   \text{terms in viscosity}.
\nonumber
\end{gather}
Analogy with the classical case but taking into account our point of view that emphasizes $C$ rather than $u$,
leads to
\begin{gather}
P = F^{-2}( p - \la \nabla_\mu C^\mu),
\nonumber
\end{gather}
where $\la$ is another viscosity coefficient.
Putting these constructions together, we obtain the following stress-energy tensor for a viscous
fluid:
\begin{align}
\begin{split}
T_{\al \be} & = (p + \varrho) u_\al u_\be + p g_{\al\be} - \la \pi_{\al\be} \nabla_\mu C^\mu 
\nonumber \\
& - \mu ( \widetilde{\nabla}_\al \widetilde{C}_\be +
  \widetilde{\nabla}_\be \widetilde{C}_\al +
\widetilde{C}^\la(  \widetilde{\nabla}_\la \widetilde{C}_\al \widetilde{C}_\be + 
  \widetilde{\nabla}_\la \widetilde{C}_\be \widetilde{C}_\al  ) ).
\end{split}
\end{align}
Writing $  \widetilde{\nabla}$ in terms of $\nabla$ and $\widetilde{C}$ in terms of $C$, one then obtains
(\ref{Lichnerowicz_stress}) with $\hc = 0$ and  
\begin{gather}
\la = \bv - \frac{2}{3} \sv \,\text{ and } \, \mu = \sv,
\label{lambda_mu}
\end{gather}
where the parametrization (\ref{lambda_mu}) is chosen so that one obtains the correct non-relativistic
limit. The addition of the terms in $\hc$ is now chosen in order to preserve the symmetry of $T_{\al\be}$,
the non-relativistic limit, and taking into account our philosophy of treating $C$ as fundamental.

Lichnerowicz proposed (\ref{Lichnerowicz_stress}) in 1955 \cite{LichnerowiczBookGR}, after Eckart
 postulated (\ref{Eckart_stress}) \cite{EckartViscous}, but
before the acausality of Eckart's theory had been discovered \cite{Hiscock_Lindblom_instability_1985}.
In hindsight, however, one can provide an alternative motivation for (\ref{Lichnerowicz_stress}), one that
does take the works \cite{EckartViscous} and \cite{Hiscock_Lindblom_instability_1985} into account, as follows.

As pointed out in section \ref{section_theories}, Eckart motivated (\ref{Eckart_stress}) from basic
principles of relativity and thermodynamics. Since Eckart's theory is not causal,
one may ask whether it is possible to modify (\ref{Eckart_stress}) preserving as much as possible of 
Eckart's well-motivated original argument, but in a way to solve the causality problem. Because
there is no question as to what the correct stress-energy tensor for perfect fluids is, one may further
consider to modify only the dissipative contributions to (\ref{Eckart_stress}). On the other hand,
the discussion of sections \ref{section_velocity_fluid} and \ref{section_enthalpy_current} suggests
that perhaps $C$, and not $u$, should be considered as a fundamental variable associated with the fluid
flow. One can then speculate if the lack of causality of Eckart's formulation is not at least in part
due to the failure of recognizing the importance of $C$, and therefore postulate a stress-energy 
of the same form as (\ref{Eckart_stress}), but replacing $u$ by $C$ in the viscous terms. Doing so
immediately leads to (\ref{Lichnerowicz_stress}), except for the last term 
$\sv \pi_{\al\be} u^\mu \nabla_\mu F$. One can decide to work with (\ref{Lichnerowicz_stress}) modulo
such term, as it was indeed suggested by Pichon \cite{PichonViscous}, or one can add this term
based on the idea that it measures the transport of the enthalpy (whose importance has already been 
highlighted) along the flow when viscosity is present. We notice that whether or not to include
the term $\sv \pi_{\al\be} u^\mu \nabla_\mu F$ is something that we do not need to decide here, since 
we are investigating the case where $\sv = 0$.

We conclude this section pointing out that there is one further reason to consider the argument
of the previous paragraph. As mentioned in section \ref{section_theories}, despite its lack 
of causality, Eckart's theory has been extensively used in the construction of models of relativistic 
viscosity. One such example was the work of Duez et al. \cite{DuezetallEinsteinNavierStokes},
who numerically solved Einstein's equations coupled to (\ref{Eckart_stress}), obtaining 
interesting results for the dynamics of neutron stars. Their results remain,
to the best of our knowledge, the most thorough numerical treatment of 
relativistic viscous fluids. It is sensible, therefore, to expect that, despite its limitations, (\ref{Eckart_stress})
encodes some of the important properties of fluids with viscosity, and, thus, the correct $T_{\al\be}$
should be in some respect close to (\ref{Eckart_stress}). This is exactly the case for 
(\ref{Lichnerowicz_stress}), since, as mentioned, $C$ differs from $u$ by terms of order $1/c^2$.

\section{The Einstein-Navier-Stokes system and its causal properties \label{section_ENS_causal}}

In this section we discuss some of the causality properties of the Einstein-Navier-Stokes system. 
We shall focus on the relatively simple situation of zero vorticity and an equation of state
given by (\ref{equation_state}). An assumption about the dependence of $\zeta$ on the other
thermodynamical variables will also be needed, and in order to avoid further technicalities 
we assume it to be constant. Explicitly:

\begin{assumption}
Throughout section \ref{section_ENS_causal}, assume that the fluid has zero vorticity, i.e., 
\begin{gather}
\Om_{\al\be} = 0,
\label{zero_Omega}
\end{gather}
that the equation of state is given by (\ref{equation_state}), and that $\zeta$ is constant
(but not zero).
\label{assumption_causal}
\end{assumption}

\begin{remark}
The case of non-constant $\zeta$ can be treated  with techniques similar to the ones presented here,
 but it would involve
further differentiation of the equations in order to obtain a system with the type of 
quasi-linear structure for which proposition \ref{proposition_causal} of the Appendix applies (provided also that 
suitable hypotheses are imposed on the functional form of $\zeta$; see (\ref{zeta_function})).
A more general equation of state could also be employed, in which case
ideas similar to those of \cite{DisconziViscousFluidsNonlinearity} should be invoked.
We avoid such technicalities for the sake of brevity and clarity of exposition.
Relaxing the assumption of eq. (\ref{zero_Omega}), however, is more delicate. 
Although it is likely that the case of non-zero vorticity could be handled if we are willing
to make somewhat severe restrictions on the initial data, as done in 
\cite{DisconziCzubakNonzero}, obtaining a sufficiently general causality result for Lichnerowicz's formulation
in the presence of vorticity remains a challenge.
In any case, our main goal in this section is not so much as to the generality of the results as to the possibility
of formulating a first order theory of relativistic viscous fluids that is causal in simple, albeit already
interesting, situations\footnote{Many important results in the classical theory
of fluids have been proven first for irrotational fluids,  with sometimes a gap of many years 
until generalizations that allow the inclusion of vorticity could be carried 
out. See, e.g., \cite{LannesWaterWavesBook, MajdaBertozziBookVorticity} and references therein.}.
\end{remark}

\begin{remark}
The assumption of zero vorticity is not very restrictive for many cosmological 
applications that deal with the universe at later times (see, e.g., Refs. \cite{ChrisMalik,Peebles_LS_book} and references therein).
\end{remark}

\subsection{Causality of the irrotational system\label{section_causality}}
From (\ref{Lichnerowicz_stress_bulk}), one readily computes
\begin{gather}
T_\al^\al = 3p - \varrho - 3 \zeta (F \nabla_\mu u^\mu + u^\mu \nabla_\mu F),
\nonumber
\end{gather}
so that Einstein's equations (\ref{Einstein_eq}) read
\begin{gather}
R_{\al\be} = (p+\varrho) u_\al u_\be + \frac{1}{2}(\varrho - p + 2\La) + \frac{3}{2} \zeta
(F \nabla_\mu u^\mu + u^\mu \nabla_\mu F).
\label{Einstein_eq_explicit}
\end{gather}

Throughout we shall assume that we are working in harmonic coordinates
or wave gauge. Recall that in these coordinates the Ricci tensor is given by (see, e.g., 
\cite{WaldBookGR1984})
\begin{gather}
R_{\al\be} = -\frac{1}{2} g^{\mu\nu} \partial_{\mu\nu} g_{\al\be} 
+ B_{\al\be}(\partial g),
\nonumber
\end{gather}
so that Einstein's equations (\ref{Einstein_eq_explicit}) read
\begin{gather}
g^{\mu\nu} \partial_{\mu\nu} g_{\al\be} + 3 \zeta g^{\mu\nu} u_\mu  \partial_\nu F 
= B_{\al\be}( \partial u, \partial g, \varrho, F),
\label{Einstein_simple}
\end{gather}
where we have adopted the following.
\begin{notation}
The letter $B$, with indices attached when appropriate, will be used to designate
a general expression depending on a maximum number of derivatives 
(denoted by the symbol $\partial$) of the variables indicated in its
arguments. For example, the right hand side of 
(\ref{Einstein_simple}) indicates an expression that depends on at most first derivatives of
$u$, first derivatives of $g$, and zeroth derivatives of $\varrho$ and $F$. The same letter $B$ will
be used to designate terms that may vary among equations.  Similarly, we shall use the notation
$a(\cdot ) \partial^k$ to designate a differential operator of order $k$ whose coefficients depend on at most
the number of derivatives of the variables indicated in the arguments of $a$. For instance,
the term $g^{\mu\nu} u_\mu \partial_\nu F$ in (\ref{Einstein_simple}) can be written
as $a(u,g) \partial F$, as it is a first order differential operator acting on $F$ whose coefficients
depend on at most zeroth derivatives of $u$ and $g$. The same letter $a$ will also be employed
for different expressions among the equations. It can be verified from the ensuing calculations
that  $B$ and $a$ only involve analytic expressions of their arguments (quotients, products, radicals, etc.).
Although their particular form will not be important, only the number of derivatives involved
will matter. In expressions (\ref{rho_eq_simple}) and (\ref{F_eq_simple}) below, however,
 the specific form of the coefficient
$a$ is important and so we denote it by $D$ instead.
\label{notation_lower_order}
\end{notation}

Notice that in (\ref{Einstein_simple}) we have expanded the covariant derivatives, absorbing
the Christoffel symbols (which depend on at most first derivatives of $g$) in the $B_{\al\be}$ term.
Also, we have written $u^\mu \partial_\mu F$ as $g^{\mu\nu} u_\mu \partial_\nu F$, making explicit
the dependence on the metric. Such considerations are necessary in order to write the equations
as a system of equations in $\RR^4$, as it is required by standard theorems of partial differential equations
and proposition \ref{proposition_causal} in particular.

Using  (\ref{Einstein_eq_explicit}) into (\ref{u_equation_w}) and recalling assumption \ref{assumption_causal},
we obtain
\begin{align}
\begin{split}
\zeta F g^{\mu \nu} \partial_{\mu\nu} u_\al +
a(\partial u, \partial g) \partial^2 g + a(u,g)\partial \varrho + a(\partial u, \partial g) \partial F =
B_\al(\partial u, \partial g, \varrho, F),
\end{split}
\label{u_eq_simple}
\end{align}
where we are following notation \ref{notation_lower_order}. 
We notice that the term $a(\partial u, \partial g) \partial^2 g$ comes from the expansion
of $\nabla^\mu \nabla_\mu u_\al$ in terms of partial derivatives and Christoffel symbols, which
will involve derivatives of the  Christoffel symbols and hence second derivatives of the metric.
Similarly, equations
(\ref{rho_equation_w}) and (\ref{F_equation_w}) read, respectively,
\begin{gather}
D(\partial u, \partial g, \varrho, F) g^{\mu\nu} u_\mu \partial_\nu \varrho 
= B(\partial u, \partial g, \varrho, F),
\label{rho_eq_simple}
\end{gather}
and
\begin{gather}
D(\partial u, \partial g, \varrho, F) g^{\mu\nu} u_\mu \partial_\nu F
= B(\partial u, \partial g, \varrho, F),
\label{F_eq_simple}
\end{gather}
where the coefficient $D(\partial u, \partial g, \varrho, F)$ is given explicitly by
\begin{align}
\begin{split}
D(\partial u, \partial g, \varrho, F)&  = 1 - \zeta \frac{F}{\varrho} \nabla_\mu u^\mu 
\\
&
= 1 - \zeta \frac{F}{\varrho} (g^{\mu \nu} \partial_\mu u_\nu 
+ g^\mu_\nu \Ga_{\mu \la}^\nu u^\la),
\end{split}
\label{D_coefficient}
\end{align}
where $\Ga_{\al\be}^\la$ are the Christoffel symbols.

Equations (\ref{Einstein_simple}), (\ref{u_eq_simple}), (\ref{rho_eq_simple}), and
(\ref{F_eq_simple}) give  the following 
for the irrotational Einstein-Navier-Stokes system:

\begin{align}
\left\{
\begin{matrix}
\zeta F g^{\mu \nu} \partial_{\mu\nu} u_\al  & + & a(\partial u, \partial g) \partial^2 g 
& + & a(u,g)\partial \varrho 
\\
0  & + & g^{\mu\nu} \partial_{\mu\nu} g_{\al\be} & + & 0
 \\
0  & + & 0 & + &  D(\partial u, \partial g, \varrho, F) g^{\mu\nu} u_\mu \partial_\nu \varrho
\\
0  & + & 0 & + &0  
\end{matrix}
\nonumber
\right.
\end{align}
\begin{subequations}
 \begin{alignat}{4}
 & + &&  \hspace{1cm} a(\partial u, \partial g) \partial F  
 &&  = && \hspace{0.25cm}B_\al(\partial u, \partial g, \varrho, F), 
 \label{system_u} \\
 &+ && \hspace{1cm}3 \zeta g^{\mu\nu} u_\mu  \partial_\nu F  
 &&  =  && \hspace{0.25cm} B_{\al\be}( \partial u, \partial g, \varrho, F),
  \label{system_g} \\
 & + && \hspace{1.75cm} 0 
 && = && 
 \hspace{0.25cm} B(\partial u, \partial g, \varrho, F),
 \label{system_rho} \\
 & + && \hspace{0.1cm} D(\partial u, \partial g, \varrho, F) g^{\mu\nu} u_\mu \partial_\nu F
 && = && 
 \hspace{0.25cm} B(\partial u, \partial g, \varrho, F).
 \label{system_F}
\end{alignat}
\label{system}
\end{subequations}
Notice that equation (\ref{system_u}) represents four equations corresponding to $\al=0,\dots,3$. Similarly,
(\ref{system_g}) correspond to ten equations.
We write  (\ref{system}) as
\begin{gather}
A(V,\partial)V = B(V),
\nonumber
\end{gather}
where $V=(u,g,\varrho, F)$, and  $B(V)=(B_u, B_g, B_\varrho, B_F)$ is the vector
whose entries are the right hand sides of (\ref{system}). The operator matrix $A$ is given by
\begin{align}
A(V,\partial) = \left(
\begin{matrix}
\zeta F g^{\mu \nu} \partial_{\mu\nu}   & a(\partial u, \partial g) \partial^2  
&  a(u,g)\partial & a(\partial u, \partial g) \partial 
 \\
0   & g^{\mu\nu} \partial_{\mu\nu} &  0 & \zeta g^{\mu\nu} u_\mu  \partial_\nu
 \\
0   & 0  &  D(\partial u, \partial g, \varrho, F) g^{\mu\nu} u_\mu \partial_\nu & 0 
\\
0  & 0 &  0  & D(\partial u, \partial g, \varrho, F) g^{\mu\nu} u_\mu \partial_\nu
\end{matrix}
\right).
\label{matrix_system}
\end{align}
Here, $V$, $B(V)$, and (\ref{matrix_system}) are interpreted as in equations
(\ref{system}), namely, the $u$ entry in $V$ corresponds to the four components
$(u_0,\dots, u_3)$, the $g$ entry to the ten components $(g_{00},\dots, g_{\al\be}, \dots , g_{33})$. 
$B_u$ and $B_g$ also correspond to four and ten entries, respectively. The matrix (\ref{matrix_system})
is $16 \times 16$, with $\zeta F g^{\mu \nu} \partial_{\mu\nu} $ corresponding to a diagonal
$4 \times 4$ matrix and $g^{\mu\nu} \partial_{\mu\nu} $ to a diagonal $10 \times 10$ matrix.
$ a(\partial u, g) \partial^2  $ is $4 \times 10$, $ a(u,g)\partial $ and
 $a(\partial u, \partial g) \partial $ are  $4 \times 1$, and $ \zeta g^{\mu\nu} u_\mu  \partial_\nu$
 is $10 \times 1$.  $D(\partial u, \partial g, \varrho, F) g^{\mu\nu} u_\mu \partial_\nu $ is a scalar,
 i.e., $1 \times 1$, operator. To establish the desired causality result, we shall employ techniques of 
 Leray systems, with  notation and terminology as reviewed in the appendix.
 
We claim that (\ref{system}) forms a Leray system with the following choice of indices:
\begin{align}
\begin{array}{cccc}
m_1 =2, & m_2  = 2, & m_3 = 1, & m_4  = 1 \\
n_1  = 0, & n_2 = 0, & n_3  = 0, & n_4  = 0,
\end{array}
\label{indices}
\end{align}
where $m_1 = m(u)$, $m_2 \equiv m(g)$,
$m_3 =  m(\varrho)$, $m_4 = m(F)$ (recall that the indices $m$ are
associated with the variables of the system), and 
$n_1 = n(\text{equation } (\ref{system_u}) )$, 
 $n_2 = n(\text{equation } (\ref{system_g}) ) $,
 $n_3 = n(\text{equation } (\ref{system_rho}) )$, 
$n_4 = n(\text{equation } (\ref{system_F}) ) $
(recall that the indices $n$ are associated with the equations of the system). As above,
some of these indices should be interpreted as corresponding to more than one quantity.
For example, $m_1 = m(u) = 2$ means that for each $u_\al$, $\al=0,\dots,3$, we associate
the index $m(u_\al) = 2$, and $n_1 = n(\text{equation } (\ref{system_u}) )=0$ means that for each
of the four equations represented by (\ref{system_u}) we associate the index $0$.
  
Now we verify that with the choice (\ref{indices}), system (\ref{system}) has the form
indicated in the appendix (see equation (\ref{general_system})). 
Consider equation (\ref{system_u}), for which the index $n_1$ is fixed at $n_1 = 0$.
The 
first term in (\ref{system_u}) is an operator of order $m_1 - n_1 = 2$ (we take $m_1$ because this operator acts on $u$ and $m_1 = m(u)$) whose coefficients
depend on zero derivatives of $g$ and $F$, hence on at most $m_2 -n_1 -1  \equiv m(g) - n_1 -1 = 1$
derivatives of $g$ and $m_4 -n_1 -1  \equiv m(F) - n_1 -1 = 0$ derivatives of $F$ (notice the key words
``at most:" $m_2 -n_1 -1 = 1$ means that the coefficient could depend on up to first derivatives of the metric,
but it does not mean that it must involve first derivatives, so being zeroth order in $g$ as is allowed here).

The 
second term in (\ref{system_u}) is an operator of order $m_2 - n_1 = 2$ (we take $m_2$ because this operator acts on $g$ and $m_2 = m(g)$) whose coefficients
depend on first derivatives of $u$ and $g$, hence on at most $m_1 -n_1 -1  \equiv m(u) - n_1 -1 = 1$
derivatives of $u$ and $m_2 -n_1 -1  \equiv m(g) - n_1 -1 = 1$ derivatives of $g$. In a similar fashion
we verify that the remaining operators in (\ref{matrix_system}) satisfy the index structure of a Leray
system, as do the $B$ terms on the right hand side. For instance, in equation (\ref{system_g}), for 
which $n_2 = 0$ is fixed, we see that $B_{\al\be}$ involves at most
$m_1 - n_2 - 1 \equiv m(u) - n_2 - 1 = 1$ derivatives of $u$, 
$m_2 - n_2 - 1 \equiv m(g) - n_2 - 1 = 1$ derivatives of $g$, 
$m_3 - n_2 - 1 \equiv m(\varrho) - n_2 - 1 = 0$ derivatives of $\varrho$, 
and
$m_4 - n_2 - 1 \equiv m(F) - n_2 - 1 = 0$ derivatives of $F$. 

The characteristic determinant of the system is easily computed since the matrix (\ref{matrix_system})
is upper triangular. We find
\begin{align}
\begin{split}
\det A(V,\xi) & =
\zeta^4 F^4 (D(\partial u, \partial g, \varrho, F))^2 (g^{\mu\nu} \xi_\mu \xi_\nu)^{14} 
(g^{\la\tau} u_\la \xi_\tau)^2.
\end{split}
\nonumber
\end{align}
The power $14$ comes from the fact that the two first diagonal entries in (\ref{matrix_system})
represent block matrices, as explained above. It is well known (see, e.g.,
\cite{ChoquetBruhatGRBook,Lichnerowicz_MHD_book}) that $g^{\mu\nu} \xi_\mu \xi_\nu$ and
$g^{\la\tau} u_\la \xi_\tau$
 are hyperbolic polynomials
(see  \cite{Leray_book_hyperbolic}), provided that $u$ is time-like for the metric $g$. Therefore,
if $D(\partial u, \partial g, \varrho, F)$ and $F$ are non-zero, we conclude that the characteristic
determinant of (\ref{system}) is a product of hyperbolic polynomials.
A domain of dependence for the system is then given by the intersection of the characteristic cones
associated with the different hyperbolic polynomials in $\det A(V,\xi)$, which means in this 
case $g^{\mu\nu} \xi_\mu \xi_\nu$ and $g^{\la\tau} u_\la \xi_\tau$. The intersection is exactly the light-cone
of the metric $g$ (see \cite{DisconziViscousFluidsNonlinearity} for a more explicit description
of the corresponding cones and their intersection).  Invoking proposition
\ref{proposition_causal}, we conclude the following.

\begin{claim} Consider the Einstein-Navier-Stokes system with no vorticity, an equation of state
of the form (\ref{equation_state}), and constant $\zeta$, i.e., equations (\ref{system}). If $F \neq 0$, 
$D(\partial u, \partial g, \varrho, F) \neq 0$, and $u$ is time-like for the metric $g$, then the system is causal,
with disturbances propagating at most at the speed of light. Furthermore, given suitable initial data,
the system has a well-posed Cauchy problem.
\label{claim_causality}
\end{claim}

We notice that for physical solutions of the Einstein-Navier-Stokes system, $u$ will indeed be time-like.
Also, we can assume that $F \neq 0$ since otherwise the fluid will be in thermodynamic equilibrium
and viscosity will not contribute (see section \ref{section_equilibrium}). The condition 
$D(\partial u, \partial g, \varrho, F) \neq 0$ seems to be relatively mild. Recalling (\ref{D_coefficient}),
this means
\begin{gather}
 1 - \zeta \frac{F}{\varrho} \nabla_\mu u^\mu \neq 0,
\label{condition_causality}
\end{gather}
which we can expect to be generically the case. In particular, if we consider the Cauchy problem, then, given
initial data for the system, we can choose $\zeta$ small such that (\ref{condition_causality}) holds.
In this regard, it is instructive to keep in mind the discussion at the end of section \ref{section_theories}
on how one should be cautious about applicability of the equations to systems very far from equilibrium, as it will be when $\zeta$ is very large. Furthermore, notice that under the present  circumstances, i.e., choosing $\zeta$
small, the term $ 1 - \zeta \frac{F}{\varrho} \nabla_\mu u^\mu$ will in fact be positive, and thus
we can invoke claim \ref{claim_entropy_production} to assure that entropy production is non-negative. In this
situation, we obtain a formulation where both causality and non-negative entropy production
are satisfied.

We point out the following remark on the aforementioned well-posedness of the 
system. Proposition \ref{proposition_causal} ensures well-posedness in the so-called
Gevrey spaces. (For a definition see the appendix and references therein.) While more general function spaces
are probably necessarily for applications of general relativity, we would like to stress that the main
point of claim \ref{claim_causality} is the causality of the equations.  This causal behavior 
will automatically carry over to solutions involving more general spaces of functions (such as
Sobolev spaces, for example).

We finish this section by discussing the condition $\Omega = 0$. Suppose we are given a solution 
to (\ref{F_equation_w}), (\ref{rho_equation_w}), (\ref{u_equation_w}), and (\ref{Einstein_eq}), without necessarily 
$\Omega = 0$, and that this solution is unique. From the equations of motion we can then derive the following relations,
valid under the assumption that $\zeta$ is small,
\begin{gather}
u^\la u^\mu \nabla_\la \nabla_\mu F = H,
\label{vort_prop_1}
\end{gather}
\begin{gather}
\pi^\mu_\al u^\la \nabla_\la \nabla_\mu F = f u^\mu \Om_{\mu \al} + G_\al,
\label{vort_prop_2}
\end{gather}
and
\begin{gather}
d \Omega = 0,
\label{vort_prop_3}
\end{gather}
where
\begin{align}
\begin{split}
H = (1 - \zeta \frac{F}{1+z} \frac{h}{p} )^{-1}[
\frac{1}{1+z} u^\mu \nabla_\mu (\frac{F}{p} u^\la \nabla_\la p)
- 
\nabla_\la F u^\mu \nabla_\mu ( (1 - \zeta \frac{F}{1+z} \frac{h}{p} )u^\la ) ],
\end{split}
\nonumber
\end{align}
\begin{align}
\begin{split}
G_\al = -  \pi^\mu_\al \frac{ \nabla_\mu h }{h} u^\la \nabla_\la F
- \pi^\mu_\al \nabla_\mu u^\la \nabla_\la F
- \pi_\al^\mu u^\la \nabla_\la u_\mu u^\be \nabla_\be F - g \pi^\mu_\al \frac{\nabla_\mu F}{F} 
+  \frac{g}{1+z} \pi^\mu_\al \frac{\nabla_\mu p}{p}
\end{split}
\nonumber
\end{align}
where $h = \frac{1-z}{1 - \zeta \frac{F}{p} \nabla_\mu u^\mu }$, 
$g = \frac{1+z}{\zeta} \frac{p}{h}$, $ f = \frac{g}{F} $, 
and $z = 1/w$. Notice that $h$, $g$, and $f$ can be assumed strictly positive.
Equations (\ref{vort_prop_1}) and (\ref{vort_prop_2}) can be combined to produce
\begin{gather}
u^\la \nabla_\la \nabla_\al F = - u_\al H +  f u^\mu \Om_{\mu \al} + G_\al.
\label{vort_prop_4}
\end{gather}
Consider now equations (\ref{vort_prop_1}), (\ref{vort_prop_2}), and (\ref{vort_prop_3}) as 
a system for $F$ and $\Omega$. Assume that the quantities $u_\al$, $H$, $G_\al$, $g$, and $f$ are known
and defined as above using the variables coming from the given solution to  
 (\ref{F_equation_w}), (\ref{rho_equation_w}), (\ref{u_equation_w}), and (\ref{Einstein_eq}).
 Notice that $G_\al$ satisfies
$u^\al G_\al = 0$ (which then implies $\pi_\al^\mu G_\mu = G_\al$). This 
system admits a solution with $\Omega = 0$ provided that $\Omega$ vanishes at time
zero. For, setting $\Omega = 0$ in (\ref{vort_prop_4}), the right-hand side is known and then 
(\ref{vort_prop_4}) can be used to derive all higher order derivatives of $F$ at $t=0$. This information
can then be used to construct a solution to (\ref{vort_prop_4}) (with $\Om = 0$) if the initial data is analytic. Decomposing
(\ref{vort_prop_4}) in the directions orthogonal and perpendicular to $u_\al$ then gives a solution to
(\ref{vort_prop_1}), (\ref{vort_prop_2}) (again, with $\Om = 0$),  and (\ref{vort_prop_3}), the latter
being trivially satisfied by $\Om \equiv 0$.

Next, we claim that the system (\ref{vort_prop_1}), (\ref{vort_prop_2}), and (\ref{vort_prop_3}) admits a unique solution.
To show this it suffices to consider the case when $H = 0 = G_\al$ and the initial data is also zero. But then we conclude
from (\ref{vort_prop_1}) that $F = 0$. Using this into (\ref{vort_prop_2}) yields
\begin{gather}
u^\mu \Om_{\mu \al} = 0.
\label{vort_prop_5}
\end{gather}
Taking the exterior derivative of (\ref{vort_prop_5}) and using (\ref{vort_prop_3}) produces
$\cL_u \Om_{\al\be} = 0$, which then implies $\Omega = 0$. Here, $\cL_u$ is the Lie 
derivative in the direction of $u$ and we used the well-known formula
$\cL_u \Omega = d \iota_u \Omega + \iota_u d \Omega$, where $\iota_u$ is contraction with $u_\al$.

We go back to  (\ref{vort_prop_1}), (\ref{vort_prop_2}), and (\ref{vort_prop_3}), where now $F$ and $\Omega$
are derived from a given solution to (\ref{F_equation_w}), (\ref{rho_equation_w}), (\ref{u_equation_w}), and (\ref{Einstein_eq}).
Since we have showed that (\ref{vort_prop_1}), (\ref{vort_prop_2}), and (\ref{vort_prop_3}) admits a solution 
with vanishing $\Omega$  if $\Omega = 0$ initially, and this solution is unique, we conclude that the solution 
$\Omega$ coming from  (\ref{F_equation_w}), (\ref{rho_equation_w}), (\ref{u_equation_w}), and (\ref{Einstein_eq})
also has to satisfy $\Omega =0$ if this condition holds initially, since $F$ and $\Omega$ satisfy
(\ref{vort_prop_1}), (\ref{vort_prop_2}), and (\ref{vort_prop_3}) identically (by construction). 

Notice that the above argument relies on the assumption that the original system does admit 
a unique solution. We do not know, however, if this is indeed the case, so the propagation of $\Omega = 0$
is at this point only a conditional result.

\subsection{The naive linear analysis and its limitations\label{section_naive}}

Here we discuss heuristic arguments sometimes employed
when addressing the causality/non-causality of evolution equations.

When faced with complex systems such as the Einstein-Navier-Stokes equations, it is always useful
to use some simple toy-model to gain intuition on the behavior of solutions. When it comes to 
causality versus non-causality, the two canonical examples are the wave equation
\begin{gather}
u_{tt} - u_{xx} = 0,
\label{wave}
\end{gather}
which is causal, and the heat equation,
\begin{gather}
v_{t} - v_{xx} = 0,
\label{heat}
\end{gather}
which is not causal. The causality/non-causality of these equations can be shown in many different 
ways. For instance, D'Alembert's and Duhamel's formulas for (\ref{wave}) and (\ref{heat}), respectively,
easily imply the existence of a domain of dependence for (\ref{wave}), whereas any localized change
at time $t=0$ affects the solution $v$ everywhere for any $t>0$. See, e.g., 
\cite{EvansPDE} for details.

A quick heuristic way of obtaining the causality/non-causality behavior of (\ref{wave}) and (\ref{heat}) is 
as follows. Take a fixed background solution, and consider perturbations $e^{-\omega t + i k x}$. This
leads to the usual dispersion relations for the wave and heat equations,
\begin{gather}
\omega^2 + k^2 = 0,
\nonumber
\end{gather}
and
\begin{gather}
-\omega + k^2 = 0.
\nonumber
\end{gather}
From these relations, we can compute the group velocity $v_g =\frac{d \omega}{dk}$, obtaining
that $v_g$ is bounded for (\ref{wave}), but increases linearly for (\ref{heat}). Thus, the diffusion
speed can assume arbitrary large values in the case of the heat equation since $k$ can be 
as large as we want.

When one has a non-linear equation, one can apply a similar argument for the linearization
about a particular solution. If the linearization has $\partial_t$ derivatives of order $p$ and 
$\partial_x$ derivatives of order $q$, we will obtain to highest order a dispersion relation of the form
\begin{gather}
\omega^p \pm k^q = 0.
\label{derivative_couting}
\end{gather}
The group velocity will, therefore, be an unbounded function of $k$, suggesting thus a breakdown
of causality, if $q > p$. We are led to suppose that a system of equations with more spatial than
temporal derivatives will not be causal.

In order to apply these arguments to general relativity, the crucial term is ``spatial." To deal with
Einstein equations one has to choose a gauge, thus specifying  a time-coordinate. While 
geometric concepts such as time-like and space-like are, of course, intrinsically defined,
the notion of a purely spatial partial differential operator is not, in the following sense. Consider the
covariant derivative $\nabla_\al$ and the subsequent projection onto the space orthogonal to $u$,
$\pi_{\be}^\al \nabla_\al$. If $u$ is time-like, then its orthogonal will be space-like, thus 
$\pi_{\be}^\al \nabla_\al$ is a derivative along a space-like direction. However, in a given coordinate
system $\{ x^\al \}_{\al=0}^3$ with associated coordinate derivatives $\{ \partial_\al \}_{\al=0}^3$,
$\pi_{\be}^\al \nabla_\al$ will still contain $\partial_t \equiv \partial_0$ terms unless
$\partial_t$ and $u$ are parallel, which will be the case only when the hypersurfaces
$\{t = \text{constant}\}$ are orthogonal to $u$. If $u$ is the four-velocity, in general this will
not be the case.  In fact, the coordinates $\{ x^\al \}_{\al=0}^3$
are fixed ahead of time by the wave gauge (i.e., wave coordinates), and we do not have the freedom
to chose $u$ orthogonal to constant time slices. Even for the Einstein-Euler system, $u$
will be orthogonal only in very specific cases \cite{FriRenCauchy}. In other words,
even if one were to subscribe to the idea that one could determine causality by counting
$\partial_t$ and $\partial_x$ derivatives, as in (\ref{derivative_couting}), such an idea
would become ambiguous in relativity, as $\partial_t$ and $\partial_x$ carry no time or spatial
meaning in this context. Another way of saying this is that causality is a coordinate independent 
property, so it should not be affected by how we choose $\partial_t$ and $\partial_x$.

Therefore, even though the highest order derivatives in the momentum equation of 
the Einstein-Navier-Stokes system come in the form $\pi^{\al \be}   \nabla_\al \nabla_\mu$
(because of the terms in first derivative of $C$ in (\ref{Lichnerowicz_stress_bulk})), this does not mean
that such terms contain only $\partial_i$ derivatives. Yet, there are legitimate reasons
to worry that terms of the form  $\pi^{\al \be}   \nabla_\al \nabla_\mu $ will spoil causality (that is why
we previously referred to these terms as ``dangerous"). To see this, note that 
when this term acts on the terms of $T_{\al\be}$ that contain derivatives of $u$, it will produce an operator of the form
$\pi^{\al \be} \pi^{\mu\nu}   \nabla_\al \nabla_\mu$, the second projection $\pi$ coming from the stress-energy tensor.
This is a
spatial (in the intrinsic sense) operator, making it more akin an elliptic operator rather than a hyperbolic one. The latter type of operator
is the one associated with causality, whereas elliptic operators can lead to non-locality, thus
non-causal,  properties (this can be seen in the simple
example of harmonic functions, where the mean-value property says that the value of the function at a point is determined
by a contour integral, with the contour possibly taken far away from the point). Such an ellipticity character is what 
is in fact behind Pichon's proof of the non-causality of Eckart's theory \cite{PichonViscous}. To see the problem in a simple
example, consider $\pi^{\al \be} \pi^{\mu\nu}   \nabla_\al \nabla_\mu \phi $, where $\phi$ is a scalar function, and take the Minkowski metric. 
Consider the component $\be=\nu=0$ and focus on the term with two $\partial_0$ derivatives. We find
\begin{align}
\begin{split}
\pi^{\al 0} \pi^{\mu0}   \nabla_\al \nabla_\mu \phi  & =
(g^{00} + u^0 u^0)(g^{00} + u^0 u^0 ) \partial_0^2 \phi + \dots
\\
&  = (-1 + (u^0)^2 )^2 \partial_0^2 \phi + \dots
\\
& = (u^i u_i)^2\partial_0^2 \phi + \dots,
\end{split}
\nonumber
\end{align}
after using $-1 = u^\al u_\al = u^0 u_0 + u^i u_i = -(u^0)^2 + u^i u_i$. We see that the term in $\partial_0^2 \phi $,
that should have a negative coefficient for hyperbolicity (like in the wave operator $-\partial^2_t \phi + \partial^2_x \phi$)
has a positive sign. The situation in our case is more complicated because we have a further contraction with $u$, absent in 
this example with a scalar field, and we have a system of equations that needs to be analyzed in full in order
to determine whether or not the equations are hyperbolic. But this simple discussion shows that while 
$\pi^{\al \be}   \nabla_\al \nabla_\mu $ will in general contain time derivatives, its structure
disfavors hyperbolicity.

Another limitation of the previous heuristic argument is that it takes for granted that the
analysis of the linearized equation carries over to the non-linear system. This is, of course, not true
in general. The expectation here seems to be that while one generally cannot make positive statements
about a non-linear system based solely on its linearization, such a generalization should be possible
regarding negative statements. In other words, if something bad (lack of causality) happens at the 
linearized level, things should only get worse for the non-linear system.

This is all sensible, but it is ultimately, at best, a general philosophy that can provide hints
on how to deal with each particular set of equations. The point is that one cannot know how much
of the linear behavior will be suppressed by the non-linearities (a case in point is studied, 
for example, in \cite{DisconziEbinFreeBoundary2d,DisconziEbinFreeBoundary3d}, where the particular
form of the non-linearity plays a crucial role in guaranteeing that certain smallness conditions
are propagated, which would not be the case for the linearized system).

Furthermore,  ideas based on a derivative counting for scalar equations, such as in the discussion
of (\ref{derivative_couting}), may not generalize to systems where extra degrees of freedom
may lead to a bound on the speed of propagation of disturbances. For instance, consider the system
\begin{subnumcases}{\label{example_system}}
g^{\mu\nu} \partial_{\mu\nu} u_\al  = f(\partial^3_i v),
\label{example_system_1}
\\
u^\al g^{\mu\nu} \partial_{\al\mu\nu} v  = h(u),
\label{example_system_2}
\end{subnumcases}
where $g$ is a Lorentzian metric, $u$ a four-vector, $v$ a scalar function,
 $h(u)$ a function of $u$ but not of its derivatives, and $ f(\partial^3_i v)$ is some 
 analytic function of the spatial derivatives of $v$, 
i.e., of $\partial_i v$, $i=1,2,3$, up to order three, 
where as usual $x^0 = t$. From the point of view of (\ref{derivative_couting}),
one could imagine that this system to be acausal, because in (\ref{example_system_1}) we have third order
derivatives with respect to $\partial_i$, and only second order time-derivatives. However, when 
equations (\ref{example_system_1}) and (\ref{example_system_2}) are taken together, one can show
the system (\ref{example_system}) is in fact causal if $u$ is time-like. In fact, it is a Leray 
system (see appendix \ref{appendix_background}) with indices $m(u_\al) = 2$, $n(u_\al) = 0$,
$m(v) = 4$, $n(v) = 1$, whose characteristic determinant is a product of the hyperbolic 
polynomials $g^{\mu\nu} \xi\mu \xi_\nu$ and $u^\al g^{\mu\nu} \xi_\al \xi_\mu \xi_\nu$, hence
causal (if $u$ is time-like).

In the other direction, one can also exhibit examples of system that might look casual when certain
equations are taken individually, but that ultimately have a parabolic structure. One such 
example is the free-boundary Stefan problem, where one of the equations has the structure
of a transport equation, but the entire system is parabolic. See \cite{MR2464573, MR3005325}
for details.

In summary, despite the acausality of Eckart's theory and other first order theories of relativistic fluids,
and despite a (legitimate) reason for concern about the causality of Lichnerowicz's formulation
due to the presence of the spatial operators $\pi^{\al \be}   \nabla_\al \nabla_\mu $, only a detailed
analysis of the equations derived from (\ref{Lichnerowicz_stress}) can answer questions
about its causal properties. Heuristic arguments can, at best, provide insight into directions
of research, at the risk, sometimes, of creating misconceptions about the behavior of solutions.

\part{APPLICATIONS TO COSMOLOGY\label{part_applications}}

\section{Viscosity in cosmology}

The possible role of viscosity in cosmology has been extensively explored, including
both the Eckart \cite{Avelino,Balakin,Gagnon,Hipolito1,Hipolito2,Colistete,Velten, Zimdahl} and
MIS \cite{MaartensDissipative}
approaches.  Ref. \cite{Disconzi_Kephart_Scherrer_2015} examined Lichnerowicz viscosity in cosmology, but only for the case in which
the viscous fluid itself dominates the expansion.  Here we extend that discussion to expansion in a matter or radiation dominated
background, and discuss further implications of the results.

\subsection{Modified FRW equations}

We assume a flat Friedmann-Robertson-Walker (FRW) metric (in accordances with observations), with scale factor $a$ normalized to
$a=1$ at present.  For the FRW metric, we have
\begin{gather}
\nabla_\mu u^\mu = 3\frac{\dot{a}}{a},
\label{div_u}
\end{gather}
which gives
\begin{gather}
\nabla_\mu C^\mu = \dot{F} + 3 F \frac{\dot{a}}{a}.
\label{div_C}
\end{gather}
The Friedmann equations
are
\begin{align}
\begin{split}
\dot{H} + H^2 \equiv \frac{\ddot{a}}{a} & 
= - \frac{1}{6} \Big (\rho + 3p 
- 3 \zeta \dot{F} - 9 \zeta F \frac{\dot{a}}{a} \Big )  ,
\end{split}
\label{fri_ddot}
\end{align}
where $H$ is the Hubble parameter, $H = \dot a/a$, and
\begin{equation}
H^2 = \frac{\rho}{3}.
\nonumber
\end{equation}

Now consider the evolution of the density of a viscous fluid.
For a standard inviscid fluid, the evolution of the density is given by
\begin{gather}
\label{rhoevol}
a \frac{d\rho}{da} = -3(\rho + p).
\end{gather}
If instead, the fluid is viscous, then Eq. (\ref{rhoevol}) must be replaced by
\begin{gather}
\label{rhoevol2}
a \frac{d\rho}{da} = -3(\rho + p_{eff}),
\end{gather}
with
\begin{gather}
\label{peff}
p_{eff} = p + \Pi,
\end{gather}
where $\Pi$ gives the effective change in the pressure due to the viscosity.  Eqs. (\ref{rhoevol2}) and (\ref{peff})
are not unique to the particular viscous model under discussion here.  For example, for the Eckart model,
we have
\begin{gather}
\label{def_PiEckart}
\Pi = -3\zeta H.
\end{gather}
In our model $\Pi$ can be read off from the right-hand side of Eq. (\ref{fri_ddot}):
\begin{gather}
\label{def_Pi}
 \Pi = -\zeta \dot{F} - 3\zeta  F H.
\end{gather}

Now consider the evolution of a viscous fluid characterized by an equation of state parameter $w$.  
The equations of motion imply
\begin{equation}
n = n_0 a^{-3},
\nonumber
\end{equation}
where $n_0$ is the present-day rest-mass density.  Defining the effective equation of state parameter $w_{eff}$ to be given
by $w_{eff} \equiv p_{eff}/\rho$, we find \cite{Disconzi_Kephart_Scherrer_2015}
\begin{equation}
\label{w_eff}
w_{eff} = \frac{w n_0 a^{-3} - {\zeta \dot w} - {3 H \zeta (1+w)}}
{n_0 a^{-3} - {3H \zeta (1+w)}},
\end{equation}
where the density evolution of the viscous fluid is now given by
\begin{equation}
\label{drhoda}
\frac{d\ln\rho}{d\ln a} = -3(1+w_{eff}).
\end{equation}

In order to make further progress, we need to make some sort of assumption about the behavior of $\zeta$.  Previous work
has generally assumed that $\zeta$ scales as a power of the density, so we will make the same assumption here, taking
\begin{equation}
\zeta = \zeta_0 \rho^\alpha.
\nonumber
\end{equation}
Furthermore, we note that $H$ is given by $H = \sqrt{\rho_T/3}$, where
$\rho_T$ is the total density, including the contributions
from both the viscous fluid and the background matter and radiation.
Then Eqs. (\ref{w_eff}) and (\ref{drhoda}) can be combined and
simplified to yield
\begin{equation}
\label{drhofinal}
\frac{d\ln\rho}{d\ln a} = -3(1+w)
\left(\frac{1 - (\lambda/\sqrt{3})\rho^\alpha a^3 [\dot w/(1+w)]
- 2 \lambda a^3 \rho^{\alpha}\rho_T^{1/2}}
{1 - \lambda (1+w) a^3 \rho^{\alpha} \rho_T^{1/2}}\right), 
\end{equation}
where we have combined the constants $n_0$ and $\zeta_0$ into
$\lambda \equiv \sqrt{3}\zeta_0/n_0$.

Eq. (\ref{drhofinal}) provides the information needed to describe the
cosmological evolution of a viscous fluid in our model.  We now examine the
specific cases of interest.

\subsection{Cosmological Evolution of a Dark Fluid}
Consider the possibility of a viscous dark fluid that serves as dark energy.
For simplicity, we will assume that this fluid has constant equation of state parameter $w$, so
that the second term in Eq. (\ref{drhofinal}) can be dropped.

At early times, the universe will be dominated by matter or radiation, with
density given by
\begin{equation}
\rho_B = \rho_0 a^{-3(1+w_B)}
\nonumber
\end{equation}
where the $B$ subscript refers to the dominant background fluid, with $w_B = 1/3$ for radiation and
$w_B = 0$ for matter.
Setting $\rho_T = \rho_B$ in
Eq. (\ref{drhofinal}), we obtain
\begin{equation}
\label{drhoback}
\frac{d\ln\rho}{d\ln a} = -3(1+w)
\left(\frac{1 -
2 \lambda  a^3 \rho^{\alpha} \rho_0^{1/2} a^{-3(1+w_B)/2}}
{1 - \lambda (1+w) a^3 \rho^{\alpha} \rho_0^{1/2} a^{-3(1+w_B)/2}}\right), 
\end{equation}
Then during the radiation-dominated era, we have
\begin{equation}
\label{drhorad}
\frac{d\ln\rho}{d\ln a} = -3(1+w)
\left(\frac{1 
- 2 \lambda \rho_0^{1/2} a \rho^{\alpha}}
{1 - \lambda \rho_0^{1/2}(1+w) a \rho^{\alpha} }\right), 
\end{equation}
while during the matter-dominated era,
\begin{equation}
\label{drhomat}
\frac{d\ln\rho}{d\ln a} = -3(1+w)
\left(\frac{1
- 2 \lambda \rho_0^{1/2}  a^{3/2} \rho^{\alpha} }
{1 - \lambda \rho_0^{1/2} (1+w) a^{3/2} \rho^{\alpha} }\right). 
\end{equation}

The behavior of the dark fluid during the background-dominated phase
will depend on the values of $w$ and $\alpha$.  When the viscous correction
is small,
the fluid density scales in the standard way: $\rho \propto a^{-3(1+w)}$.
In this case,
we see from
Eqs. (\ref{drhorad})-(\ref{drhomat}) that the viscous correction will
increase with scale factor as as long as
\begin{equation}
\label{alphacondition1}
\alpha < \frac{1}{3(1+w)},
\end{equation}
in the
radiation-dominated era, and
\begin{equation}
\label{alphacondition2}
\alpha < \frac{1}{2(1+w)}.
\end{equation}
in the matter-dominated era.
In each case, there is an asymptotic solution to Eqs.
(\ref{drhorad})-(\ref{drhomat}) which applies as long as the dark fluid
is subdominant, namely
\begin{eqnarray}
\label{radscaling}
\rho &\propto& a^{-1/\alpha}, ~~ w_{eff} = \frac{1}{3\alpha}-1~~  ({\rm radiation-dominated})\\
\label{matscaling}
\rho &\propto& a^{-3/2\alpha}, ~~w_{eff} = \frac{1}{2\alpha} -1 ~~({\rm matter-dominated}).
\end{eqnarray}

As long as $w_{eff} < 0$, the dark fluid will eventually come to
dominate the expansion at late times.  At this point, Eq. (\ref{drhofinal})
becomes
\begin{equation}
\label{drhodark}
\frac{d\ln\rho}{d\ln a} = -3(1+w)
\left(\frac{1
- 2 \lambda a^3 \rho^{\alpha+1/2}}
{1 - \lambda (1+w) a^3 \rho^{\alpha+1/2}}\right), 
\end{equation}
The evolution of the density given by Eq. (\ref{drhodark}) was
discussed briefly in Ref. \cite{Disconzi_Kephart_Scherrer_2015}, but
we provide a more complete set of solutions here.
As noted in Ref. \cite{Disconzi_Kephart_Scherrer_2015}, the viscous correction
dominates at late times as long as
\begin{equation}
\label{alphacondition3}
\alpha < \frac{1-w}{2(1+w)}.
\end{equation}
In that case Eq. (\ref{drhodark})
has an asymptotic solution for which $w_{eff}$ is constant, namely
\begin{equation}
\label{darkscaling}
\rho \propto a^{-3/(\alpha + 1/2)},~~w_{eff} = \frac{1-2\alpha}{1+2\alpha}.
\end{equation}

Now we must determine the stability of the solutions given by Eqs. (\ref{radscaling}), (\ref{matscaling}), and
(\ref{darkscaling}).  Note
that Eqs. (\ref{drhorad}), (\ref{drhomat}), and
(\ref{drhodark}) are all of the form
\begin{equation}
\frac{d\ln\rho}{d\ln a} = -3(1+w)
\left(\frac{1
- 2 A a^\beta \rho^\gamma}
{1 - A (1+w) a^\beta \rho^\gamma} \right), 
\end{equation}
where $A$, $\beta$, and $\gamma$ are constants.  The solutions given by Eqs. (\ref{radscaling}), (\ref{matscaling}), and
(\ref{darkscaling}) can then be written as $\rho =  \rho_1 a^{-\beta/\gamma}$.  Writing the perturbed solution
in the form $\rho = \rho_1 a^{-\beta/\gamma}(1+\epsilon)$ and expanding out to linear order in $\epsilon$, we obtain
\begin{equation}
\frac{d\ln \epsilon}{d \ln a} = \frac{\beta}{\gamma} A \rho_1 \left(\frac{1-w}{[1-2A \rho_1][1 - (1+w)A\rho_1)]}\right). 
\end{equation}
Thus, the condition for stability is $\beta/\gamma < 0$.  For the solutions in the matter or radiation dominated eras
(Eqs. \ref{radscaling}-\ref{matscaling}), this corresponds to $\alpha < 0$, while for the background-dominated case
(Eq. \ref{darkscaling}) stability requires $\alpha < -1/2$.

While the solutions given in Eqs. (\ref{radscaling}), (\ref{matscaling}), (\ref{darkscaling})
apply only when $\alpha < 0$ (for the background-dominated case) or when $\alpha < -1/2$ (for the dark
fluid dominated case), these are not the only parameter ranges resulting in interesting deviation
from the inviscid dark fluid case.  Instead, as long as Eqs. (\ref{alphacondition1}), (\ref{alphacondition2}),
and (\ref{alphacondition3}) are satisfied in the radiation, matter, and dark fluid dominated regimes,
respectively, $w_{eff}$ will inevitably evolve below $-1$, driving the dark fluid to
phantom-like behavior, and driving the universe to a future big rip singularity.  For example,
for a viscosity that is independent of the density ($\alpha = 0$), we find this behavior for all values
of $w$ (assuming $-1 \le w \le 1$), while for a viscosity  that is linearly-dependent on density ($\alpha =1$) we
need $1+w < 1/3, ~1/2, ~2/3$ in the radiation, matter, and dark fluid dominated regimes, respectively.

To illustrate this behavior, we have numerically integrated Eq. (\ref{drhofinal}) for a model
with nonrelativistic matter and a dark fluid
with a constant value of $w=-0.9$.  We define the present ($a=1$) to correspond to $\Omega_{DE} = 0.7$, and
we then choose $\lambda$ in Eq. (\ref{drhofinal}) to yield $w_{eff} = -1.1$ at the present.  [Recall
that $w_{eff}$ is related to the left-hand side of Eq. (\ref{drhofinal}) via Eq. (\ref{drhoda})].  In
Fig. 1, we show the evolution of $w_{eff}$ as a function of the scale factor for $\alpha = 0,1,2$.
Fig. 1 clearly illustrates the crossing of the phantom divide at $w_{eff} = -1$, and the evolution
of $w_{eff}$ is quite distinct compared to its evolution in most other models for dark energy.  However, we also
see that the evolution of $w_{eff}$ is nearly independent of the value of $\alpha$,
i.e., the scaling of $\zeta$ with $\rho$.  This is easy to understand in terms of the way we have derived
our results.  By considering models in which $w_{eff}$ evolves only between $-0.9$ and $-1.1$,
we automatically limit ourselves to the case in which $\rho_{DE}$ evolves very slowly with the
scale factor.  Consequently, the factor $\rho^{\alpha}$ in Eq. (\ref{drhofinal}) is nearly constant
regardless of the value of $\alpha$.  We would not expect this result to hold for models in which $w_{eff}$ deviates
more strongly from $-1$.
\begin{figure}[t]
\centerline{\epsfxsize=3.7truein\epsfbox{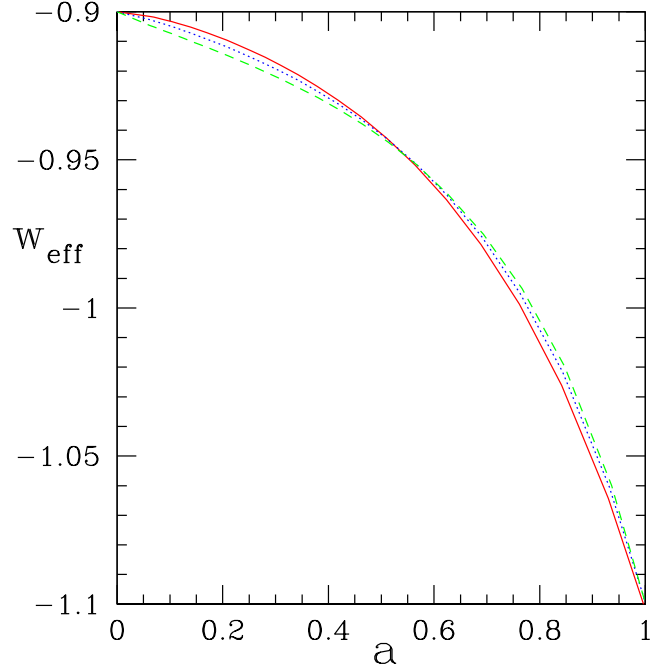}}
\caption{Evolution of the effective equation of state parameter $w_{eff}$ 
as a function of the scale factor $a$, for a viscous dark fluid
with $w=-0.9$, where the present ($a=1$) corresponds to
$\Omega_{DE} = 0.7$, and the viscosity scales as $\zeta = \zeta_0 \rho^\alpha$ with
$\alpha = 0$ (solid, red), $\alpha = 1$ (blue, dotted), $\alpha = 2$ (green, dashed). In all three cases,
$\zeta_0$ is chosen to give $w_{eff} = -1.1$ at the present.}
\end{figure}

One of the most interesting possibilities to emerge from the study of
cosmological viscosity is that a viscous fluid with a dustlike equation
of state ($w=0$) might serve as a unified model for dark matter and
dark energy.  This is a natural possibility to explore for
the case of Eckart viscosity; in this model viscosity is subdominant
at early times, so the viscous fluid behaves like dark matter,
while at late times the viscosity drives $w_{eff}$ to
$-1$ asymptotically, giving an accelerated expansion for the universe.
(See, e.g., Refs. \cite{Avelino,Hipolito1}).  However, Li and Barrow
\cite{LiBarrow} and Velten and Schwarz \cite{VeltenSchwarz} showed
that such models face severe observational difficulties.

It is therefore natural to ask whether the viscous model discussed here
can evade the difficulties with models based on Eckart viscosity.  Our
model gives different evolution for a $w=0$ fluid than does the Eckart model.
As in the Eckart model, a single fluid with $w=0$ evolves, at late times,
toward a fluid that can drive the accelerated expansion, but in our case
the late-time behavior gives $w_{eff} < -1$ instead of $w_{eff} = -1$.  However, the results
in Refs. \cite{LiBarrow,VeltenSchwarz} depend crucially
on the effects of the assumed model for viscosity on perturbation growth, so it
is not straightforward to apply them directly to our model.  This
question is currently under investigation.

Our model can result in a particularly interesting variant of this idea for the appropriate choice of
$\alpha$.  Consider the case where $\alpha$ lies in the range
\begin{equation}
\label{specialmodel}
\frac{1}{3(1+w)} < \alpha < \frac{1}{2(1+w)}.
\end{equation}
Then, from Eqs. (\ref{alphacondition1})-(\ref{alphacondition2}), we see that the viscous correction
to the equation of state parameter for the fluid will decay during the radiation-dominated era,
but increase during the matter dominated era.  This offers the possibility of generating models
in which the onset of dark energy domination is triggered by the transition from the radiation-dominated
era to the matter-dominated era, helping to resolve the coincidence problem.  If
we take $w=0$, the dark fluid will behave like matter during the radiation-dominated era, but then
viscous corrections will drive it toward dark energy behavior
after it becomes the dominant component.  Note that Eq.
(\ref{specialmodel}) violates the stability condition in the background-dominated case, but
this just means that the evolution of the density will not be given by Eqs. (\ref{radscaling})-(\ref{matscaling}).
However, as noted above, the viscosity correction will nonetheless increase with time, driving an accelerated
expansion and ultimately a future big rip.

As we mentioned earlier, when $w=-1$, the Lichnerowicz
bulk viscosity is identically zero (see also the discussion in Ref. 
\cite{Disconzi_Kephart_Scherrer_2015}), although 
the viscous contributions reappear when $w$ is below $-1$. This is
a curious feature of this model for two reasons. First,
 it is a remarkable coincidence that in Lichnerowicz's formulation, where viscosity plays an important role in crossing the phantom divide, the viscous contributions vanish at the value 
 $w=-1$, which in turn corresponds precisely to the value when a perfect fluid crosses the phantom barrier. Whether this is a mere coincidence or a hint of some deeper feature warrants further investigation. Second, the value $w=-1$ is special in the perfect fluid case as it is widely used
 to model
 a cosmological constant. Since  Lichnerowicz's model reduces to a perfect fluid when 
 $w=-1$, the analyses performed in the former case apply verbatim to our case as well.


\appendix

\section{Mathematical background\label{appendix_background}}

Our goal in this appendix is to state proposition \ref{proposition_causal} below,
which will be used in section \ref{section_ENS_causal}
 to establish the causality of the Einstein-Navier-Stokes system.
We begin by fixing notation. 
We assume familiarity with standard terminology of hyperbolic differential 
equations as explained, for example, in \cite{Courant_and_Hilbert_book_2}.
Let $X = \RR \times  \RR^n$, and denote coordinates on $X$ by
$\{ x^\al \}_{\al = 0}^n$, thinking of $x^0 \equiv t$ as the time-variable.
By  $\partial^k$ we shall denote any $k^{\text{th}}$ order derivative. 
We use the notation $a = a(x,\partial^k)$, $x \in X$ for  
a linear differential operator of order $k$. We can write
\begin{gather}
a(x,\partial^k) = \sum_{ |\al | \leq k } a_\al(x) \partial^\al,
\nonumber
\end{gather}
where $\al = (\al_0, \al_1, \al_2, \dots, \al_n)$ is a multi-index
and
\begin{gather}
\partial^\al \equiv \partial^{|\al |}_{i_0 i_1  i_2 \cdots i_n} \equiv
\frac{\partial^{|\al |}}{\partial x_0^{\al_0} \partial x_1^{\al_1} \partial x_2^{\al_2} \cdots \partial x_n^{\al_n} } 
\equiv \partial^{\al_0}_{x^0} \partial^{\al_1}_{x^1} \partial^{\al_2}_{x^2} \cdots \partial^{\al_n}_{x^n},
\nonumber
\end{gather}
where   $|\al | = \al_0 +  \al_1 + \al_2 + \cdots + \al_n$.

To a given linear differential operator of order $k$ we can associate, at each point $x \in X$ 
and for each co-vector $\xi$ at $x$,
 a polynomial of order $k$ in the variable $\xi$
 obtained by replacing the derivatives by $\xi$. More precisely, for each 
$k^{\text{th}}$ order derivative of $a(x,\partial^k)$, i.e., 
\begin{gather}
\partial^{|\al |}_{i_0 i_1  i_2 \cdots i_n},
\nonumber
\end{gather}
$|\al | = k$,
we associate the monomial
\begin{gather}
\xi^\al \equiv \xi_0^{\al_0} \xi_1^{\al_1} \xi_2^{\al_2} \cdots \xi_n^{\al_n},
\nonumber
\end{gather}
where $\xi = (\xi_0, \xi_1,\xi_2,\dots, \xi_n)$, forming in this way the polynomial
\begin{gather}
a(x,\xi) = \sum_{ |\al | = k } a_\al(x) \xi^\al.
\nonumber
\end{gather}
In the above we only take account of the highest order derivatives of the operator, i.e., 
derivatives of order $k$.

Consider a system of $N$ partial differential 
equations and $N$ unknowns in $X$,
and
denote the unknown as 
$V=(v^I)$, $I=1,\dots, N$. 
The system is a called a \textbf{Leray system} \cite{Leray_book_hyperbolic}
if it is possible 
to attach to each unknown $v^I$ a non-negative integer $m_I$,  and to
each equation a non-negative integer $n_J$, such that the system reads
\begin{gather}
h^J_I(x,\partial^{m_K - n_J -1} v^K, \partial^{m_I - n_J}) v^I
+ b^J(x, \partial^{m_K - n_J - 1} v^K) = 0, \, J=1, \dots, N.
\label{general_system}
\end{gather}
Here,
$h^J_I(x,\partial^{m_K - n_J -1} v^K, \partial^{m_I - n_J})$
is a homogeneous differential operator of order $m_I - n_J$ (which can
be zero), whose coefficients depend on at most
$m_K - n_J -1$ derivatives of $v^K$, $K=1,\dots N$. The remaining terms, 
$b^J(x,\partial^{m_K - n_J - 1} v^K)$, also depend on at most 
$m_K - n_J -1$ derivatives of $v^K$,  $K=1,\dots N$. 
If $m_K - n_J < 0$, then the corresponding term in the $J^{\text{th}}$
equation does not depend on $v^K$.
These indices, sometimes called \textbf{Leray indices}, are defined only up to an
overall additive constant.

Since the differential operators $h^J_I(x,\partial^{m_K - n_J -1} v^K, \partial^{m_I - n_J})$ 
in (\ref{general_system}) are allowed to depend on the unknown, in general they will not be linear.
Given a sufficiently regular $V$,   the
\begin{gather}
h^J_I(x,\partial^{m_K - n_J -1} v^K, \partial^{m_I - n_J})
\nonumber
\end{gather}
are well-defined linear
 operators, and we can ask about their hyperbolicity properties.
The \textbf{characteristic determinant}  of (\ref{general_system}) 
at $x \in X$ and \emph{for a given} $V$
 is the polynomial in the co-vector $\xi$ given by 
\begin{gather}
p(x,V, \xi) = \det(h^J_I(x, \partial^{m_K - n_J -1} v^K(x), \xi)).
\label{characteristic_det}
\end{gather}
It is not difficult to see that $p$ is a homogeneous polynomial in the variable $\xi$ of degree 
\begin{gather}
 \ell \equiv \sum_{I=1}^N m_I - \sum_{J=1}^N n_J.
\nonumber
\end{gather}
Two features of Leray systems, and its corresponding characteristic determinant (\ref{characteristic_det}),
should be highlighted. First, equations of different orders are 
allowed in the system. In fact, the only constraint to the order of the differential operators
is that indices $m_I$ and $n_J$ can be chosen as indicated above. On the other hand,
notice that it is not always possible to choose such indices so that the system
looks like (\ref{general_system}), which simply means that not every system of differential
equations is a Leray system.
Second, from the form (\ref{general_system}) it follows that a certain unknown
$v^I$ may have its highest number of derivatives appearing in the terms $b^J$ with $J\neq I$, 
which, in turn, do not enter in the characteristic determinant (which, as discussed below, is the quantity
determining the causality properties of the system). For example, suppose that the first 
unknown $v^1$ and the first equation $J=1$ are, respectively, the metric $g_{\al\be}$ and 
Einstein's equations written in harmonic gauge:
\begin{gather}
-\frac{1}{2} g^{\mu\nu} \partial_{\mu \nu} g_{\al\be} +
b_{\al\be}(x, \partial g, \partial^{m_K - n_1} \varphi^K) = 0,
\label{example_Einstein}
\end{gather}
where $\varphi^K$, $k=2,\dots, N$ represent the matter fields. Suppose further that the indices
are such that $m_1 = 4$, which necessarily implies $n_1 = 2$ since 
(\ref{example_Einstein}) is a second order equation for the unknown $v^1 \equiv g_{\al\be}$.
Suppose also that the equation for $\varphi^2$ is first order and $m_2 = 1$, which then
requires $n_2 = 0$. From (\ref{general_system}) if follows that the term $b^2$ can depend
on at most $m_1 - n_2-1$ derivatives of $v^1$, i.e., it can involve \emph{three}
derivatives of $g_{\al\be}$, more derivatives than appear in the equation for $g_{\al\be}$ 
itself\footnote{Obviously, since we are talking about a system, the same unknown figures in several equations,
but it is customary to consider each equation as the main equation for a given unknown. For instance,
when we study Einstein's equations coupled to matter, the metric appears in all the equations, including those
of the matter fields (say, Maxwell or Euler equations), but we treat Einstein's equations
themselves as the ``equations for the metric."}, i.e.,
in (\ref{example_Einstein}).

\begin{remark}
One of the important insights of Leray in his seminal work \cite{Leray_book_hyperbolic}
was to realize that the $b^J$ terms, which one usually considers as lower order terms,
can in fact depend on higher derivatives of some of the unknowns $v^I$ when $I \neq J$. 
How many more derivatives 
are allowed is dictated by the indices $m_I$ and $n_J$. At the same time, we see that
when $I=J$, i.e., when we consider the main equation\footnote{See previous footnote.} for the 
unknown $v^I$, we see from 
the form of $b^I$ in (\ref{general_system}) that the maximum number of derivatives
of $v^I$ will necessarily be in the principal part $h^I_J \equiv h^I_I$ of the equation, agreeing with
the usual notion that the principal part, which gives the characteristic determinant,
contains the top number of derivatives. For a concrete example of this situation where the highest
order derivatives of some unknowns are in the $b^J$ terms, see the equations of magneto-hydrodynamics
coupled to Einstein's equations in section 45 of \cite{Lichnerowicz_MHD_book}.
\end{remark}

Consider the Cauchy problem for (\ref{general_system}), with Cauchy data  given 
on the an initial surface $\Si = \{ x^0 = 0 \}$.
Assume that when $V$ is replaced by the Cauchy data, 
the characteristic determinant (\ref{characteristic_det}) is a product of $s$
hyperbolic polynomials,
\begin{gather}
p(x,V,\xi) = p_{1}(x,\xi) \cdots p_s(x,\xi).
\label{product_diagonalization}
\end{gather}
Let $\ell_q$ be the order of $p_q(x,\xi)$, and suppose that
\begin{gather}
\max_q \ell_q \geq \max_I m_I - \min_J n_J.
\label{condition_diagonalization_theorem} 
\end{gather}
Building on the works of Leray \cite{Leray_book_hyperbolic} and Leray-Ohya
\cite{LerayOhyaLinear, LerayOhyaNonlinear}, Choquet-Bruhat showed
that (\ref{general_system}) is well-posed and causal when (\ref{condition_diagonalization_theorem})
holds. We summarize the result as:

\begin{proposition} Consider the Cauchy problem for (\ref{general_system}).
Assume that the characteristic determinant $p(x,V,\xi)$, with $V$ replaced by the Cauchy data,
is a product of $s$ hyperbolic polynomials,
as in (\ref{product_diagonalization}), and that (\ref{condition_diagonalization_theorem}) holds.
 Then, (\ref{general_system}) has a unique solution in a suitable Gevrey class and in
some neighborhood $ X^\prime = [0,T^\prime] \times \RR^n$ of $\Si = \{ x^0 = 0 \}$.
 Furthermore, the system enjoys the finite propagation speed
property, with the domain of dependence given by the intersection of the cones determined
by the hyperbolic polynomials $p_q(x,\xi)$, $x \in X^\prime$.
\label{proposition_causal}
\end{proposition}

We refer the reader to the above works of Leray, Leray-Ohya, and Choquet-Bruhat 
for a precise mathematical statement and proof of the proposition. Variants of
the statement of proposition \ref{proposition_causal} can also be found (without proofs)
in \cite{ChoquetBruhatGRBook, DisconziCzubakNonzero, DisconziViscousFluidsNonlinearity,
Lichnerowicz_MHD_book}. See also \cite{ChoquetetallAnalysisManifolds,Courant_and_Hilbert_book_2} for more background on Leray's theory of hyperbolic equations.

Gevrey spaces,
originally introduced by Gevrey in \cite{GevreyOriginal},
consist of a special class of smooth functions.
They have been extensively used in fluid dynamics (see, e.g.,
\cite{TadmorBesovGevrey,TitiGevreyNavier,TitiGevreyParabolic,TemamGevrey})
and hyperbolic partial differential equations (see, e.g., \cite{MizohataCauchyProblem})
(the main type of differential equations relevant in the study of Einstein's equations), 
and they are not completely foreign to problems in general relativity \cite{ChoquetBruhatGRBook,Lichnerowicz_MHD_book}.
We refer the reader to \cite{RodinoGevreyBook} for an introduction to Gevrey spaces.

Although applications to general relativity usually require larger classes 
of functions than Gevrey spaces (see, e.g., \cite{KlainermanNicoloBook,RingstromCauchyBook}),
the relevant conclusion of proposition \ref{proposition_causal} for our purposes is the causality
of the equations. The domains of dependence and influence of the system
are given by the intersection of the interior of the cones determined by the
hyperbolic polynomials in (\ref{product_diagonalization}). In general relativity, we are interested
in the case where such intersection coincides with the light-cone. In particular, because
of the finite speed of propagation property, proposition \ref{proposition_causal} can be adapted
to manifolds via a standard argument in local charts.

{\bf Acknowledgments}
MMD is partially supported by NSF grant 1305705.
  The work of TWK and RJS is supported by US DOE grant DE-SC0010504.

\bibliographystyle{plain}
\bibliography{References.bib}

\end{document}